\documentclass[11pt,a4paper]{amsart}
\usepackage[margin=2.5cm]{geometry}

\usepackage{amsmath,amsfonts,amssymb,amscd,amsthm}
\usepackage[usenames,dvipsnames,svgnames]{xcolor}

\usepackage[utf8x]{inputenc}
\usepackage{graphicx}
\graphicspath{{figures/}}
\usepackage{caption}
\usepackage{subcaption}
\usepackage[numbers]{natbib}
\usepackage{doi}
\usepackage{autonum}
\usepackage{makecell}
\usepackage{hyperref}
\usepackage{acronym}
\usepackage{listings}
\usepackage{textgreek}
\usepackage[frozencache,cachedir=.]{minted}

\hypersetup{
    bookmarks=true,
    breaklinks=true,
    bookmarksopen=true,
    pdftitle={A user-guide to Gridap -grid-based approximation of partial differential equations in Julia-},    
    pdfauthor={Francesc Verdugo and Santiago Badia},     
    colorlinks=true,       
    linkcolor=black,          
    citecolor=blue,        
    filecolor=black,      
    urlcolor=blue           
}

\definecolor{bg}{rgb}{0.93,0.93,0.93}

\DeclareMathOperator{\grad}{\boldsymbol\nabla}

\acrodef{dof}[DOF]{Degree Of Freedom}
\acrodef{dg}[DG]{Discontinuous Galerkin}
\acrodef{rt}[RT]{Raviart-Thomas}
\acrodef{pde}[PDE]{Partial Differential Equation}
\acrodef{fe}[FE]{Finite Element}
\acrodefplural{fe}[FEs]{Finite Elements}
\acrodef{jit}[JIT]{just-in-time}
\acrodef{ode}[ODE]{Ordinary Differential Equation}

\usepackage{amsmath}
\usepackage{graphicx}
\usepackage{hyperref}
\usepackage{url}
\usepackage[colorinlistoftodos, textwidth=1.5\marginparwidth, shadow]{todonotes}

\usepackage{graphicx}
\usepackage{color}
\usepackage{framed}
\usepackage{verbatim}
\usepackage{fancyvrb}

\definecolor{shadecolor}{gray}{.92}
\definecolor{incolor}{rgb}{0,0,.7}
\definecolor{outcolor}{rgb}{.65,0,0}
\definecolor{syntaxcolor}{rgb}{.65,0,0}

\newcommand{\shb}[1]{\texttt{#1}}

\def\review1{{\color{blue} 1}}

\newcommand{\fig}[1]{Fig.~\ref{#1}}
\newcommand{\sect}[1]{Sect.~\ref{#1}}

\begin{document}

\def\gridap{Gridap}
\def\julia{{Julia}}
\def\paraview{Paraview}

\title[A user-guide to Gridap]{A user-guide to Gridap --Grid-based approximation of partial differential equations in Julia--}

\author[F. Verdugo]{Francesc Verdugo$^{1,*}$}

\author[S. Badia]{Santiago Badia$^{1,2}$}

\thanks{\null\\
$^1$ Centre Internacional de M\`etodes Num\`erics en Enginyeria, Esteve Terrades 5, E-08860 Castelldefels, Spain.\\
$^2$ School of Mathematics, Monash University, Clayton, Victoria, 3800, Australia.\\
$^*$ Corresponding author.\\
E-mails: {\tt fverdugo@cimne.upc.edu} (FV), {\tt santiago.badia@monash.edu} (SB)}

\date{\today}

\begin{abstract}
  We present \gridap{}, a new scientific software library for the numerical approximation of partial differential equations (PDEs) using grid-based approximations. {\gridap} is an open-source software project exclusively written in the Julia programming language. The main motivation behind the development of this library is to provide an easy-to-use framework for the development of complex PDE solvers in a dynamically typed style without sacrificing the performance of statically typed languages. This work is a tutorial-driven user guide to the library. It covers some popular linear and nonlinear PDE systems for scalar and vector fields, single and multi-field problems, conforming and nonconforming finite element discretizations, on structured and unstructured meshes of simplices and hexahedra. %

\end{abstract}

\maketitle




\section{Introduction} \label{sec:int}

Many phenomena around us can be modeled by \acp{pde}. These equations cannot be solved analytically in general and numerical methods, e.g., \acp{fe} \cite{Johnson2009} or finite volumes \cite{Versteeg2007}, have been developed to approximate these phenomena. These methods discretize the original \ac{pde} and end up with discrete (non)linear systems that can be solved exploiting computing resources. The computational cost required to compute accurate simulations increases as problems become more complex and larger scales are considered. The toughest problems, e.g., in plasma physics or turbulent flows, demand the largest supercomputing resources or are still unfeasible.

The development of high-performance scientific software for the numerical approximation of \acp{pde} that can effectively exploit increasing computational resources is a key research area with a broad impact in advanced scientific and engineering applications. Such codes are usually written in static programming languages, mainly C/C++ and Fortran 95/03/08. This static knowledge of types allows compilers to perform optimizations and generate high-performance machine code. On the downside, these languages are also related to poor code productivity.

On the other side, dynamic languages like Python or MATLAB allow one to write high-level codes. It results in much more expressive implementations that require less lines of code to implement a given algorithm since no type declarations are needed. The compilation and link times are eliminated, boosting productivity in interactive IDEs, a standard REPL (real-eval-print loops), or Jupyter notebooks. The price of productivity is performance. Data types must be checked at runtime, generating much slower codes.


Different types of combinations of these two worlds have been proposed to enjoy both performance and productivity in \ac{pde} simulation software. Vectorization can be used to speed up dynamic language codes, e.g., using Numpy with Python, thus transferring the computationally intensive blocks to pre-compiled C or Fortran libraries. On the other hand, C++ scientific software libraries, e.g., deal.ii \cite{Bangerth2007} or FEniCS \cite{alnaes_fenics_2015}, provide high-level Python interfaces. Application experts whose needs are already covered by an existing \ac{pde} library can readily use these high-level interfaces to enjoy advanced simulation tools. However, such approach is not satisfactory when users need functionalities not provided by the low-level library. This is known as the two programming language problem.

Probably, the most advanced scientific libraries for \ac{pde} approximations have been developed in academic/research environments. In research teams on computational science and computational or numerical mathematics, PhD students and postdocs want to explore the next numerical discretization scheme, or to test variations of existing algorithms and methods. There are two different approaches. The first one involves to develop poor-performance ad-hoc high-level prototypes that prevent their usage in advanced applications. The second option is to master a low-level library with a steep learning curve and poor productivity.

Julia \citep{Bezanson2017} is a new language that has being designed having computationally intensive numerical algorithms in mind, as Fortran decades ago. Julia aims to combine the performance of statically-typed programming languages with the productivity of dynamically-type ones. As a result, there is no need to use low-level codes or vectorization for getting performance, thus eliminating the two-language problem. Data types are not required to be specified at all instances but flow through the program through an automatic type inference system. If types can be inferred by the Julia \ac{jit} compiler, performance is comparable to the one of static languages. In any case, the design of performant Julia code is not obvious, and developers must become familiar with new concepts, e.g., type-stability \cite{TheJuliaProject}.

Julia is not an object-oriented language as C++ or Fortran 2003/08. It uses a multiple dispatch paradigm and many concepts of functional programming (like higher order functions and closures) but permits mutable structures for numerical performance, e.g., array computations. The use of packages in Julia is very easy, which is also essential for productivity. On top of all this, Julia provides an excellent package handler, built-in documentation generators, or straightforward unit testing programming tools.

The Julia ecosystem has some high-quality libraries exclusively written in Julia for optimization \cite{jump_web}, \acp{ode} \cite{diffeqsjl_web}, or data science \cite{flux_web}. However, in our opinion, there were not high quality libraries for grid-based approximation of \acp{pde} back in January 2019. The existing libraries \cite{juliafem_web} were restricted to specific techniques, applications, or discretization orders. \gridap{} started as a project to generate performant code while keeping a high coding productivity. It is an open-source software project hosted at github \cite{gridap_repo} exclusively written in the \julia{} programming language. Our aim is to reduce as much as possible the set up time for new users and developers and even use the library for educational purposes in computational mathematics units.

The design of \gridap{} is quite unique for a \ac{fe} software package since it is not a simple translation of an existing C/C++ or FORTRAN code. In contrast, the library makes use of lazy-data structures that represent objects (e.g., elemental matrices and vectors) on the entire computational domain. This allows us to hide assembly loops and other core computations from the user-code leading to a very compact, user-friendly, syntax.


In any case, it is not the aim of this work to provide a detailed presentation of the design patters being used at \gridap{} or provide details related to its internal structures. Instead, this work is a tutorial-based introduction to \gridap{} through a set of tutorials solving some common \ac{pde} systems with conforming \ac{fe} and \ac{dg} methods. Our objective is to show that \gridap{} provides very powerful abstractions that make simple the implementation of \ac{pde} approximations. Even though the project itself is very young, \gridap{} already provides different types of conforming \ac{fe} methods, e.g., nodal Lagrangian \acp{fe} for grad-conforming approximations (e.g., linear elasticity or thermal analysis) or non-nodal \acp{fe} (e.g., Raviart-Thomas spaces for div-conforming approximations of flow in porous media). Using the previously described integration machinery, one can also implement \ac{dg} schemes. The way differential forms are written in Julia resembles their statement in mathematical notation, making the implementation of new drivers very straightforward. The treatment of multi-field problems is also very natural.

In Sect. \ref{sec:installing_gridap} we provide some instruction about how to install and run \gridap{}. A set of tutorials in increasing order of complexity are described in Sects. \ref{sec:poisson} to \ref{sec:ins}. Along the tutorials, we present the main abstractions a user should familiarize with to implement their own drivers and how these abstractions are built and combined in some cases of interest. We draw some conclusions in Sect. \ref{sec:con}.

\section{Gridap installation} \label{sec:installing_gridap}

\gridap{} is a registered package  in the official \julia{} package index.  Therefore, its installation is straight-forward via the \julia{} package manager. To install \gridap{}, first install a recent version of \julia{} (at least version 1.0), e.g., by downloading the binaries from the \julia{} project webpage \cite{julialang_web}.  Once \julia{} is available in your system, \gridap{} can be installed as any other registered \julia{} package: open the \julia{} REPL (i.e., execute the \julia{} binary), type \texttt{]} to enter package mode, and install the package as follows
\begin{minted}[bgcolor=bg]{julia}
pkg> add Gridap
\end{minted}
That's all. The other \julia{} packages used in the tutorials bellow (e.g., LineSearches) can be installed in the same way. The code provided in this paper is for \gridap{} version 0.8.0 or higher. To see the specific version of \gridap{} you have installed, open a \julia{} REPL, type \texttt{]} to enter package mode, and check the status of the package as follows
\begin{minted}[bgcolor=bg]{julia}
pkg> status Gridap
\end{minted}
If you have installed \gridap{} in the past and you want to update to the most recent version, use the command
\begin{minted}[bgcolor=bg]{julia}
pkg> update Gridap
\end{minted}
For more information about how to deal with \julia{} packages, see the official documentation of the \julia{} package manager \cite{pkg_web}.
 
 Each tutorial below provides a set of code snippets, with all the code needed to solve the underlying \ac{pde}.  Note that the code snippets have to be executed in the same order as they appear in the text. The code in one tutorial is self-contained and independent from the others. It is recommended to not mix the code of different tutorials in the same \julia{} session to avoid conflicts.
 
An extended version of the tutorials presented in this paper can be found as Jupyter notebooks and html pages in the github repository \url{https://github.com/gridap/Tutorials}. In the code below, we import \ac{fe} meshes stored in \shb{json} files. To access these files, clone this repository, and navigate to the \shb{models} folder. The required \shb{json} files are hosted there.
\begin{minted}[bgcolor=bg]{shell}
$ git clone https://github.com/gridap/Tutorials.git
$ cd Tutorials/models
\end{minted}

\newcounter{tutorial}
\newcommand{\tutorial}{Tutorial~\arabic{tutorial}}

\stepcounter{tutorial}
\section{Poisson equation (\tutorial)} \label{sec:poisson}

In this tutorial, we will learn
\begin{itemize}
  \item  How to solve a simple \ac{pde} in Julia with \gridap{}
  \item  How to load a discrete model (aka a \ac{fe} mesh) from a file
  \item  How to build a conforming Lagrangian \ac{fe} space
  \item  How to define the different terms in a weak form
  \item  How to impose Dirichlet and Neumann boundary conditions
  \item  How to visualize results
\end{itemize}

\subsection{Problem statement}
\begin{figure}[ht!]
\includegraphics[width=0.6\textwidth]{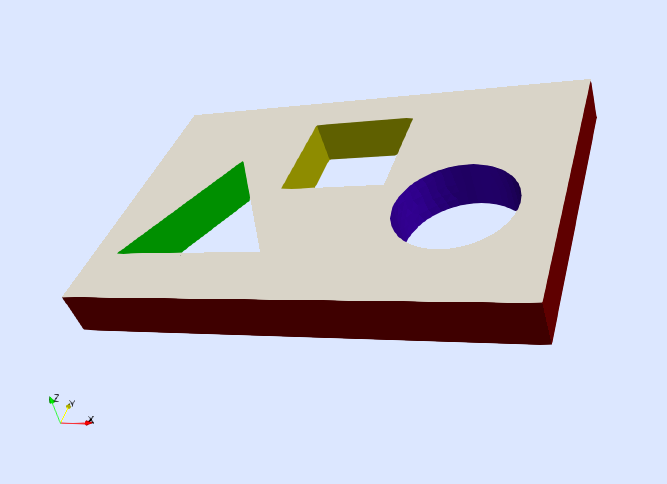}
\caption{\tutorial: View of the 3D computational domain.}
\label{fig:poisson_model}
\end{figure}

In this first tutorial, we provide an overview of a complete simulation pipeline in \gridap{}, from the construction of the \ac{fe} mesh to the visualization of the computed results. To this end, we consider a simple model problem, the Poisson equation.
 We want to solve the Poisson equation on the 3D domain depicted in \fig{fig:poisson_model} with Dirichlet and Neumann boundary conditions. Dirichlet boundary conditions are applied on $\Gamma_{\rm D}$, being the outer sides of the prism (marked in red in \fig{fig:poisson_model}). Non-homogeneous Neumann conditions are applied to the internal boundaries $\Gamma_{\rm G}$, $\Gamma_{\rm Y}$, and $\Gamma_{\rm B}$ (marked in green, yellow and blue respectively). And homogeneous Neumann boundary conditions are applied in $\Gamma_{\rm W}$, the remaining portion of the boundary (marked in white).

The problem to solve in this tutorial is: find the scalar field $u$ such that

 \begin{align}
 \left\lbrace
 \begin{aligned}
 -\Delta u = f  \ &\text{in} \ \Omega,\\
 u = g \ &\text{on}\ \Gamma_{\rm D},\\
 \nabla u\cdot n = h \ &\text{on}\  \Gamma_{\rm N},\\
 \end{aligned}
 \right.
 \end{align}
 being $n$ the outwards unit normal vector to the Neumann boundary $\Gamma_{\rm N} \doteq \Gamma_{\rm G}\cup\Gamma_{\rm Y}\cup\Gamma_{\rm B}\cup\Gamma_{\rm W}$. In this example, we chose $f(x) = 1$, $g(x) = 2$, and $h(x)=3$ on $\Gamma_{\rm G}\cup\Gamma_{\rm Y}\cup\Gamma_{\rm B}$ and $h(x)=0$ on $\Gamma_{\rm W}$. The variable $x$ is the position vector $x=(x_1,x_2,x_3)$.

 \subsection{Numerical scheme}

 To solve this \ac{pde}, we use a conventional Galerkin finite element (FE) method with conforming Lagrangian \ac{fe} spaces (see, e.g., \cite{Johnson2009} for specific details on this formulation). The corresponding weak form is: find $u\in U_g$ such that $ a(u,v) = b(v) $ for all $v\in V_0$, where $U_g$ and $V_0$ are the subset of functions in $H^1(\Omega)$ that fulfill the Dirichlet boundary condition $g$ and $0$ respectively. The bilinear and linear forms for this problems are
 \begin{align}
   a(u,v) \doteq \int_{\Omega} \nabla v \cdot \nabla u \ {\rm d}\Omega, \quad b(v) \doteq \int_{\Omega} v\ f  \ {\rm  d}\Omega + \int_{\Gamma_{\rm N}} v\ h \ {\rm d}\Gamma_{\rm N}.
 \end{align}
The problem is solved numerically by approximating the spaces $U_g$ and $V_0$ by their discrete counterparts associated with a \ac{fe} mesh of the computational domain $\Omega$. As we have anticipated, we consider standard conforming Lagrangian \ac{fe} spaces for this purpose.

The implementation of this numerical scheme in \gridap{} is done in a user-friendly way thanks to the abstractions provided by the library. As it will be seen below, all the mathematical objects involved in the definition of the discrete weak problem have a correspondent representation in the code.

 \subsection{Setup}

 The first step is to load the \gridap{} library in the current \julia{} session. If you have configured your \julia{} environment properly (see \sect{sec:installing_gridap}), it is simply done with the line
\begin{minted}[bgcolor=bg]{julia1}
using Gridap
\end{minted}

\subsection{Discrete model}

As in any FE simulation, we need a discretization of the computational domain (i.e., a FE mesh). All geometrical data needed for solving a FE problem is provided in \gridap{} by types inheriting from the abstract type \shb{DiscreteModel}. In the following line, we build an instance of \shb{DiscreteModel} by loading a \shb{json} file.
\begin{minted}[bgcolor=bg]{julia1}
model = DiscreteModelFromFile("model.json")
\end{minted}
The file \shb{model.json} is a regular \shb{json} file that includes a set of fields that describe the discrete model. It was generated using the GMSH mesh generator \cite{Geuzaine2009} together with the {GridapGmsh} package \cite{gridapgmhs_web}. The two main steps used to generate the  \shb{model.json} file are: First, we create a \shb{model.msh} file with {GMSH} (which contains a \ac{fe} mesh and information about user-defined physical boundaries in {GMSH} format). Then, this file is converted to the \gridap{}-compatible \shb{model.json} file using the conversion tools available in the {GridapGmsh} package. See the documentation of the {GridapGmsh} project \cite{gridapgmhs_web} for particular details.
You can easily inspect the discrete model in \paraview{} \cite{Ayachit_2015} by writing it in \shb{vtk} format.
\begin{minted}[bgcolor=bg]{julia1}
writevtk(model,"model")
\end{minted}

The previous line generates four different files \shb{model\_0.vtu}, \shb{model\_1.vtu}, \shb{model\_2.vtu}, and \shb{model\_3.vtu} containing the vertices, edges, faces, and cells present in the discrete model. Moreover, you can easily visualize which boundaries are defined on the geometrical objects of the model.

\begin{figure}[ht!]
\includegraphics[width=0.6\textwidth]{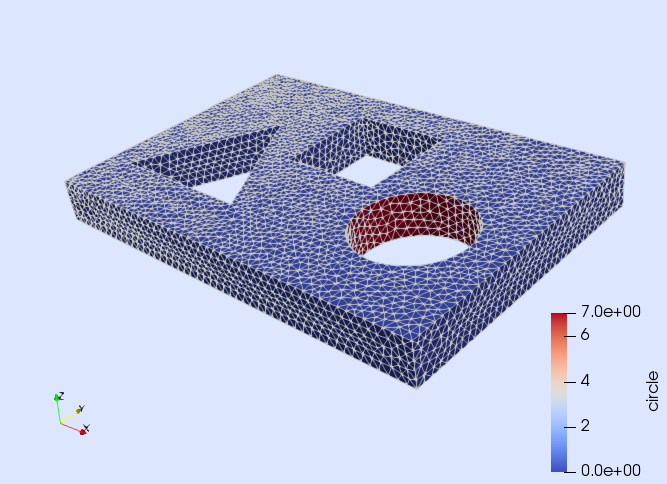}
\caption{\tutorial: View of the faces in the discrete model. Faces on the circular perforation are identified with the ``circle" tag.}
\label{fig:poisson_faces_on_circle}
\end{figure}

For instance, if you want to see which faces of the model are on the boundary $\Gamma_{\rm B}$ (i.e., the walls of the circular perforation), open the file \shb{model\_2.vtu} and chose coloring by the element field ``circle". You should see that only the faces on the circular perforation have a value different from zero (see \fig{fig:poisson_faces_on_circle}).

\begin{figure}[ht!]
\includegraphics[width=0.6\textwidth]{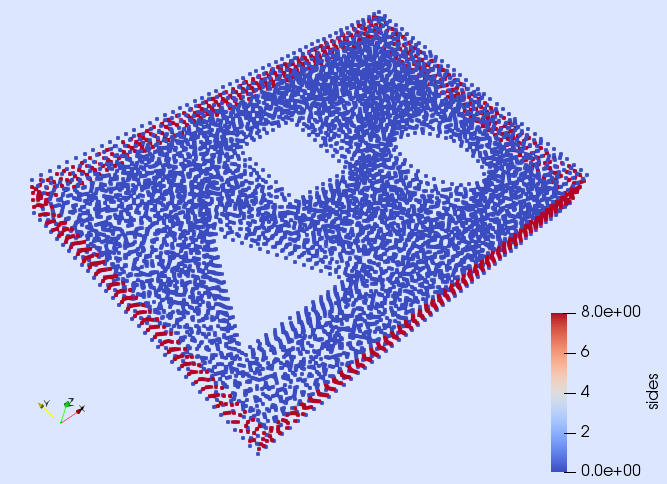}
\caption{\tutorial: View of the vertices in the discrete model. Vertices on the outer sides of the model are identified with the ``sides" tag.}
\label{fig:poisson_vertices_on_sides}
\end{figure}

It is also possible to see which vertices are on the Dirichlet boundary $\Gamma_{\rm D}$. To do so, open the file \shb{model\_0.vtu} and chose coloring by the field "sides" (see \fig{fig:poisson_vertices_on_sides}).

That is, the boundary $\Gamma_{\rm B}$ (i.e., the walls on the circular perforation) is identified within the model with the ``circle" tag and the Dirichlet boundary $\Gamma_{\rm D}$ with the ``sides" tag. In addition, the walls of the triangular perforation $\Gamma_{\rm G}$ and the walls of the square perforation $\Gamma_{\rm Y}$ are identified in the model with the names ``triangle" and ``square" respectively. You can easily check this by opening the corresponding file in \paraview{}.

\subsection{FE spaces}

Once we have a discretization of the computational domain, the next step is to generate a discrete approximation of the \ac{fe} spaces $V_0$ and $U_g$ (i.e. the test and trial FE spaces) of the problem. To do so, first, we are going to build a discretization of $V_0$ as the standard Conforming Lagrangian FE space (with zero boundary conditions) associated with the discretization of the computational domain. The approximation of the FE space $V_0$ is built as follows:

\begin{minted}[bgcolor=bg]{julia1}
V0 = TestFESpace(
  reffe=:Lagrangian, order=1, valuetype=Float64,
  conformity=:H1, model=model, dirichlet_tags="sides")
\end{minted}
Here, we have used the \shb{TestFESpace} constructor, which constructs a particular FE space (to be used as a test space) from a set of options described as key-word arguments. With the options \shb{reffe=:Lagrangian}, \shb{order=1}, and  \shb{valuetype=Float64}, we define the local interpolation at the reference \ac{fe} element. In this case, we select a scalar-valued, first order, Lagrangian interpolation. In particular, the value of the shape functions will be represented with  64-bit floating point numbers. With the key-word argument \shb{conformity} we define the regularity of the interpolation at the boundaries of the cells in the mesh. Here, we use \shb{conformity=:H1}, which means that the resulting interpolation space is a subset of $H^1(\Omega)$ (i.e., continuous shape functions). On the other hand, with the key-word argument \shb{model}, we select the discrete model on top of which we want to construct the \ac{fe} space. Finally, we pass the identifiers of the Dirichlet boundary via the \shb{dirichlet\_tags} argument. In this case, we mark as Dirichlet all objects of the discrete model identified with the \shb{"sides"} tag. Since this is a test space, the corresponding shape functions will vanish on these geometrical objects.

Once the space $V_0$ is discretized, we proceed with the approximation of the trial space $U_g$.
\begin{minted}[bgcolor=bg]{julia1}
g(x) = 2.0
Ug = TrialFESpace(V0,g)
\end{minted}
To this end, we have used the \shb{TrialFESpace} constructors. Note that we have passed a function representing the value of the Dirichlet boundary condition, when building the trial space. That is, functions in this \ac{fe} space will be equal to $g$ on the Dirichlet boundary.

\subsection{Numerical integration}

Once we have built the interpolation spaces, the next step is to set up the machinery to perform the integrals in the weak form numerically. Here, we need to compute integrals on the interior of the domain $\Omega$ and on the Neumann boundary $\Gamma_{\rm N}$. In both cases, we need two main ingredients. We need to define an integration mesh (i.e. a triangulation of the integration domain), plus a Gauss-like quadrature in each of the cells in the triangulation. In \gridap{}, integration meshes are represented by types inheriting from the abstract type \shb{Triangulation}. For integrating on the domain $\Omega$, we build the following triangulation and quadrature:

\begin{minted}[bgcolor=bg]{julia1}
trian = Triangulation(model)
degree = 2
quad = CellQuadrature(trian,degree)
\end{minted}
Here, we build a triangulation from the cells of the discrete model and define a quadrature of degree  2 in the cells of this triangulation. This is enough for integrating the corresponding terms of the weak form exactly for an interpolation of order 1.

On the other hand, we need a special type of triangulation, represented by the type	\linebreak \shb{BoundaryTriangulation}, to integrate on the boundary. Essentially, a  \shb{BoundaryTriangulation} is a particular type of \shb{Triangulation} that is aware of which cells in the model are touched by faces on the boundary. We build an instance of this type from the discrete model and the names used to identify the Neumann boundary as follows:

\begin{minted}[bgcolor=bg]{julia1}
neumanntags = ["circle", "triangle", "square"]
btrian = BoundaryTriangulation(model,neumanntags)
bquad = CellQuadrature(btrian,degree)
\end{minted}
In addition, we have created a quadrature of degree 2 on top of the cells in the triangulation for the Neumann boundary.

\subsection{Weak form}

With all the ingredients presented so far, we are ready to define the weak form. This is done by means of types inheriting from the abstract type \shb{FETerm}. In this tutorial, we will use the sub-types \shb{AffineFETerm} and \shb{FESource}. An \shb{AffineFETerm} is a term that contributes both to the system matrix and the right-hand-side vector, whereas a \shb{FESource} only contributes to the right hand side vector. Here, we use an \shb{AffineFETerm} to represent all the terms in the weak form that are integrated over the interior of the domain $\Omega$.
\begin{minted}[bgcolor=bg]{julia1}
f(x) = 1.0
a(u,v) = ∇(v)*∇(u)
b_Ω(v) = v*f
t_Ω = AffineFETerm(a,b_Ω,trian,quad)
\end{minted}

In the first argument of the \shb{AffineFETerm} constructor, we pass a function that represents the integrand of the bilinear form $a(\cdot,\cdot)$. The second argument is a function that represents the integrand of the part of the linear form $b(\cdot)$ that is integrated over the domain $\Omega$. The third argument is the \shb{Triangulation} on which we want to perform the integration (in that case the integration mesh for $\Omega$), and the last argument is the \shb{CellQuadrature} needed to perform the integration numerically. Since the contribution of the Neumann boundary condition is integrated over a different domain, it cannot be included in the previous \shb{AffineFETerm}. To account for it, we use a \shb{FESource}:

\begin{minted}[bgcolor=bg]{julia1}
h(x) = 3.0
b_Γ(v) = v*h
t_Γ = FESource(b_Γ,btrian,bquad)
\end{minted}
In the first argument of the \shb{FESource} constructor, we pass a function representing the integrand of the Neumann boundary condition. In the two last arguments we pass the triangulation and quadrature for the Neumann boundary.


 \subsection{FE Problem} At this point, we can build the FE problem that, once solved, will provide the numerical solution we are looking for. A FE problem is represented in \gridap{} by types inheriting from the abstract type \shb{FEOperator} (both for linear and nonlinear cases). Since we want to solve a linear problem, we use the concrete type \shb{AffineFEOperator}, i.e., a problem represented by a matrix and a right hand side vector.
\begin{minted}[bgcolor=bg]{julia1}
op = AffineFEOperator(Ug,V0,t_Ω,t_Γ)
\end{minted}
Note that the \shb{AffineFEOperator} object representing our \ac{fe} problem is built from the trial and test \ac{fe} spaces \shb{V0} and \shb{Ug}, and the objects \shb{t\_Ω} and \shb{t\_Γ} representing the weak form.

 \subsection{Solver phase}

\begin{figure}[ht!]
\includegraphics[width=0.6\textwidth]{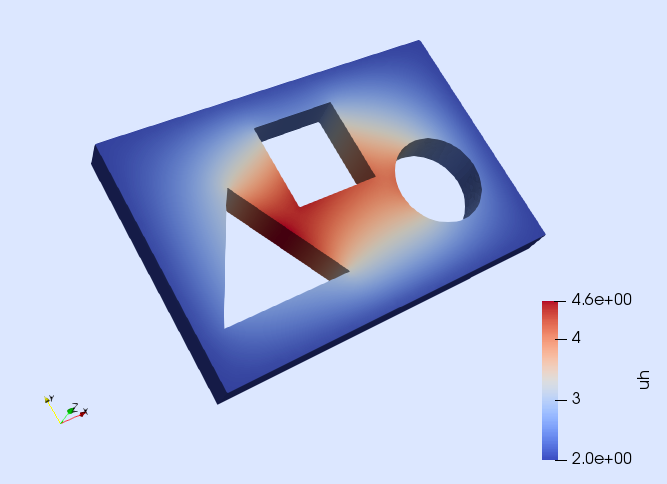}
\caption{\tutorial: View of the computed numerical solution.}
\label{fig:poisson_uh}
\end{figure}

 We have constructed a FE problem, the last step is to solve it. In \gridap{}, FE problems are solved with types inheriting from the abstract type \shb{FESolver}. Since this is a linear problem, we use a \shb{LinearFESolver}:

\begin{minted}[bgcolor=bg]{julia1}
ls = LUSolver()
solver = LinearFESolver(ls)
\end{minted}
 Note that \shb{LinearFESolver} objects are built from a given algebraic linear solver. In this case, we use a LU factorization. Now we are ready to solve the \ac{fe} problem with the \ac{fe} solver as follows:
\begin{minted}[bgcolor=bg]{julia1}
uh = solve(solver,op)
\end{minted}
The \shb{solve} function returns the computed numerical solution \shb{uh}. This object is an instance of \shb{FEFunction}, the type used to represent a function in a FE space. We can inspect the result by writing it into a \shb{vtk} file:
\begin{minted}[bgcolor=bg]{julia1}
writevtk(trian,"results",cellfields=["uh"=>uh])
\end{minted}
 which will generate the file \shb{results.vtu} having a nodal field named \shb{"uh"} containing the solution of our problem (see \fig{fig:poisson_uh}). 

\stepcounter{tutorial}
\section{Linear elasticity (\tutorial)} \label{sec:elasticity}

In this tutorial, we will learn
\begin{itemize}
  \item How to approximate vector-valued problems
  \item How to solve problems with complex constitutive laws
  \item How to impose Dirichlet boundary conditions only in selected components
  \item How to impose Dirichlet boundary conditions described by more than one function
\end{itemize}

\subsection{Problem statement}

\begin{figure}[ht!]
\includegraphics[width=0.6\textwidth]{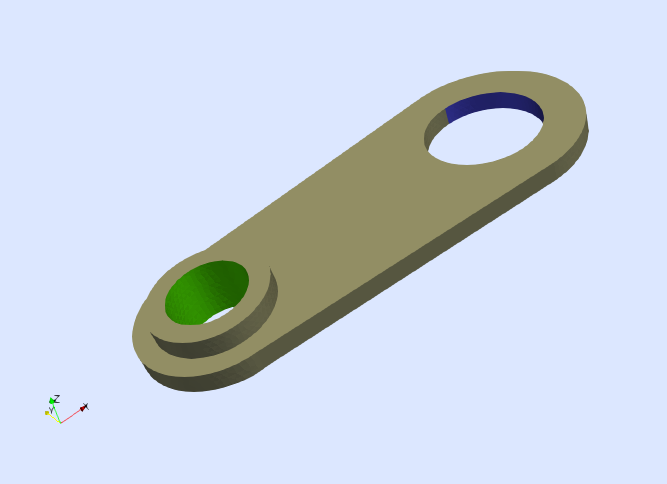}
\caption{\tutorial: View of the 3D computational domain.}
\label{fig:elasticity_solid}
\end{figure}

In this tutorial, we detail how to solve a linear elasticity problem defined on the 3D domain depicted in \fig{fig:elasticity_solid}.

We impose the following boundary conditions. All components of the displacement vector  are constrained to zero on the surface $\Gamma_{\rm G}$, which is marked in green in \fig{fig:elasticity_solid}. On the other hand, the first component of the displacement vector is prescribed to the value $\delta\doteq 5$mm on the surface $\Gamma_{\rm B}$, which is marked in blue in \fig{fig:elasticity_solid}. No body or surface forces are included in this example. Formally, the \ac{pde} to solve is
\begin{align}
\left\lbrace
\begin{aligned}
-∇\cdot\sigma(u) = 0 \ &\text{in} \ \Omega,\\
u = 0 \ &\text{on}\ \Gamma_{\rm G},\\
u_1 = \delta \ &\text{on}\ \Gamma_{\rm B},\\
\sigma(u)\cdot n = 0 \ &\text{on}\  \Gamma_{\rm N}.\\
\end{aligned}
\right.
\end{align}

The variable $u$ stands for the unknown displacement vector, the vector $n$ is the unit outward normal to the Neumann boundary $\Gamma_{\rm N}\doteq\partial\Omega\setminus\left(\Gamma_{\rm B}\cup\Gamma_{\rm G}\right)$ and $\sigma(u)$ is the stress tensor defined as
\begin{equation}
\sigma(u) \doteq \lambda\ {\rm tr}(\varepsilon(u)) \ I +2 \mu \  \varepsilon(u),
\end{equation}
where $I$ is the 2nd order identity tensor, and $\lambda$ and $\mu$ are the \emph{Lamé parameters} of the material. The operator $\varepsilon(u)\doteq\frac{1}{2}\left(\nabla u + (\nabla u)^t \right)$ is the symmetric gradient operator (i.e., the strain tensor). Here, we consider material parameters corresponding to aluminum with Young's modulus $E=70\cdot 10^9$ Pa and Poisson's ratio $\nu=0.33$. From these values, the Lamé parameters are obtained as $\lambda = (E\nu)/((1+\nu)(1-2\nu))$ and $\mu=E/(2(1+\nu))$.

\subsection{Numerical scheme}

As in previous tutorial, we use a conventional Galerkin \ac{fe} method with conforming Lagrangian \ac{fe} spaces. For this formulation, the weak form is: find $u\in U$ such that $ a(u,v) = 0 $ for all $v\in V_0$, where $U$ is the subset of functions in $V\doteq[H^1(\Omega)]^3$ that fulfill the Dirichlet boundary conditions of the problem, whereas $V_0$ are functions in $V$ fulfilling $v=0$ on $\Gamma_{\rm G}$ and $v_1=0$ on $\Gamma_{\rm B}$. The bilinear form of the problem is
\begin{equation}
a(u,v)\doteq \int_{\Omega} \varepsilon(v) : \sigma(u) \ {\rm d}\Omega.
\end{equation}

The main differences with respect to previous tutorial is that we need to deal with a vector-valued problem, we need to impose different prescribed values on the Dirichlet boundary, and the integrand of the bilinear form $a(\cdot,\cdot)$ is more complex as it involves the symmetric gradient operator and the stress tensor. However, the implementation of this numerical scheme is still done in a user-friendly way since all these features can be easily accounted for with the abstractions in the library.

\subsection{Discrete model}

\begin{figure}[ht!]
\includegraphics[width=0.6\textwidth]{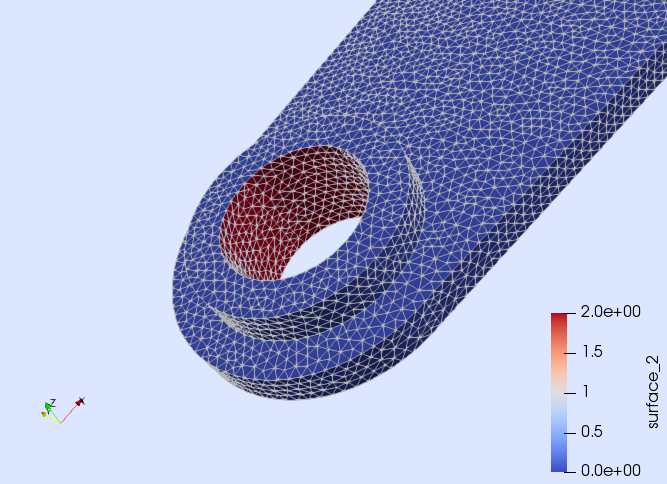}
\caption{\tutorial: Close-up view of the faces in the discrete model.}
\label{fig:elasticity_solid-surf2}
\end{figure}

We start by loading the discrete model from a file
\begin{minted}[bgcolor=bg]{julia1}
using Gridap
model = DiscreteModelFromFile("solid.json")
\end{minted}
In order to inspect it, we write the model to vtk
\begin{minted}[bgcolor=bg]{julia1}
writevtk(model,"model")
\end{minted}
and open the resulting files with \paraview{} (see \fig{fig:elasticity_solid-surf2}). The boundaries $\Gamma_{\rm B}$ and $\Gamma_{\rm G}$ are identified  with the names \shb{"surface\_1"} and \shb{"surface\_2"} respectively.  For instance, if you visualize the faces of the model and color them by the field \shb{"surface\_2"} (see \fig{fig:elasticity_solid-surf2}), you will see that only the faces on $\Gamma_{\rm G}$ have a value different from zero.

\subsection{Vector-valued FE space}

Here, we need to build a vector-valued \ac{fe} space, which is done as follows:
\begin{minted}[bgcolor=bg]{julia1}
order = 1
V0 = TestFESpace(
  reffe=:Lagrangian, order=order, valuetype=VectorValue{3,Float64},
  conformity=:H1, model=model, dirichlet_tags=["surface_1","surface_2"],
  dirichlet_masks=[(true,false,false), (true,true,true)])
\end{minted}
As in previous tutorial, we construct a continuous Lagrangian interpolation of order 1. The vector-valued interpolation is selected via the option \shb{valuetype=VectorValue\{3,Float64\}}, where we use the type \shb{VectorValue\{3,Float64\}}, which is the way \gridap{} represents vectors of three \shb{Float64} components. We mark as Dirichlet the objects identified with the tags \shb{"surface\_1"} and \shb{"surface\_2"} using the \shb{dirichlet\_tags} argument. Finally, we chose which components of the displacement are actually constrained on the Dirichlet boundary via the \shb{dirichlet\_masks} argument. Note that we constrain only the first component on the boundary $\Gamma_{\rm B}$ (identified as \shb{"surface\_1"}), whereas we constrain all components on $\Gamma_{\rm G}$ (identified as \shb{"surface\_2"}).

The construction of the trial space is slightly different in this case. The Dirichlet boundary conditions are described with two different functions, one for boundary $\Gamma_{\rm B}$ and another one for $\Gamma_{\rm G}$. These functions can be defined as
\begin{minted}[bgcolor=bg]{julia1}
g1(x) = VectorValue(0.005,0.0,0.0)
g2(x) = VectorValue(0.0,0.0,0.0)
\end{minted}
From functions \shb{g1} and \shb{g2}, we define the trial space as follows:
\begin{minted}[bgcolor=bg]{julia1}
U = TrialFESpace(V0,[g1,g2])
\end{minted}
Note that the functions \shb{g1} and \shb{g2} are passed to the \shb{TrialFESpace} constructor in the same order as the boundary identifiers are passed previously in the \shb{dirichlet\_tags} argument of the \shb{TestFESpace} constructor.

\subsection{Constitutive law}

Once the FE spaces are defined, the next step is to define the weak form.  In this example, the construction of the weak form requires more work than in previous tutorial since we need to account for the constitutive law that relates strain and stress. In this case, the integrand of the bilinear form of the problem is written in the code as follows:
\begin{minted}[bgcolor=bg]{julia1}
a(u,v) = inner( ε(v), σ(ε(u)) )
\end{minted}
The symmetric gradient operator is represented by the function \shb{ε} provided by \gridap{} (also available as \shb{symmetric\_gradient}). However, function \shb{σ} representing the stress tensor is not predefined in the library and it has to be defined ad-hoc by the user. The way function \shb{σ} and other types of constitutive laws are defined  in \gridap{} is by using the supplied macro \shb{@law}:

\begin{minted}[bgcolor=bg]{julia1}
using LinearAlgebra: tr
const E = 70.0e9
const ν = 0.33
const λ = (E*ν)/((1+ν)*(1-2*ν))
const μ = E/(2*(1+ν))
@law σ(ε) = λ*tr(ε)*one(ε) + 2*μ*ε
\end{minted}
The macro \shb{@law} is placed before a function definition.  The arguments of the function annotated with the \shb{@law} macro represent the values of different quantities at a generic integration point. In this example, the argument represents the strain tensor, from which the stress tensor is to be computed using the Lamé operator. Note that the implementation of function \shb{σ} is very close to its mathematical definition. Under the hood, the \shb{@law} macro adds an extra method to the annotated function. The newly generated method can be used as \shb{σ(ε(u))} in the definition of a bilinear form (as done above), or as \shb{σ(ε(uh))}, in order to compute the stress tensor associated with a \shb{FEFunction} object  \shb{uh}.

\subsection{Solution of the FE problem}

The remaining steps for solving the FE problem are essentially the same as in previous tutorial.  We build the triangulation and quadrature for integrating in the volume, we define the terms in the weak form, and we define the FE problem. Finally, we solve it.
\begin{minted}[bgcolor=bg]{julia1}
trian = Triangulation(model)
degree = 2*order
quad = CellQuadrature(trian,degree)
t_Ω = LinearFETerm(a,trian,quad)
op = AffineFEOperator(U,V0,t_Ω)
uh = solve(op)
\end{minted}
Note that in the construction of the \shb{AffineFEOperator} we have used a \shb{LinearFETerm} instead of an \shb{AffineFETerm} as it was done in previous tutorial. The \shb{LinearFETerm} is a particular implementation of \shb{FETerm}, which only leads to contributions to the system matrix (and not to the right hand side vector). This is what we need here since the body forces are zero. Note also that we do not have explicitly constructed a \shb{LinearFESolver}. If a \shb{LinearFESolver} is not passed to the \shb{solve} function, a default solver is created and used internally.

\begin{figure}[ht!]
\includegraphics[width=0.6\textwidth]{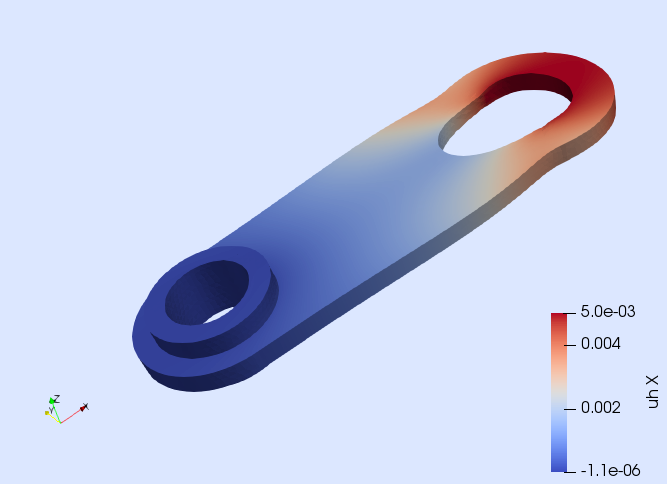}
\caption{\tutorial: View of the numerical solution (deformation magnified 40 times).}
\label{fig:elasticity_disp_ux_40}
\end{figure}

Finally, we write the results to a file. Note that we also include the strain and stress tensors into the results file.
\begin{minted}[bgcolor=bg]{julia1}
writevtk(trian,"results",cellfields=["uh"=>uh,"epsi"=>ε(uh),"sigma"=>σ(ε(uh))])
\end{minted}
We can see in \fig{fig:elasticity_disp_ux_40} that the surface  $\Gamma_{\rm B}$ is pulled in $x_1$-direction and that the solid deforms accordingly.

\stepcounter{tutorial}
\section{$p$-Laplacian equation (\tutorial)} \label{sec:tutorial02}

In this tutorial, we will learn
\begin{itemize}
  \item How to solve a simple nonlinear \ac{pde} in \gridap{}
  \item How to define the weak residual and its Jacobian
  \item How to setup and use a nonlinear solver
  \item How to define new boundaries from a given discrete model
\end{itemize}

\subsection{Problem statement}

\begin{figure}[ht!]
\includegraphics[width=0.6\textwidth]{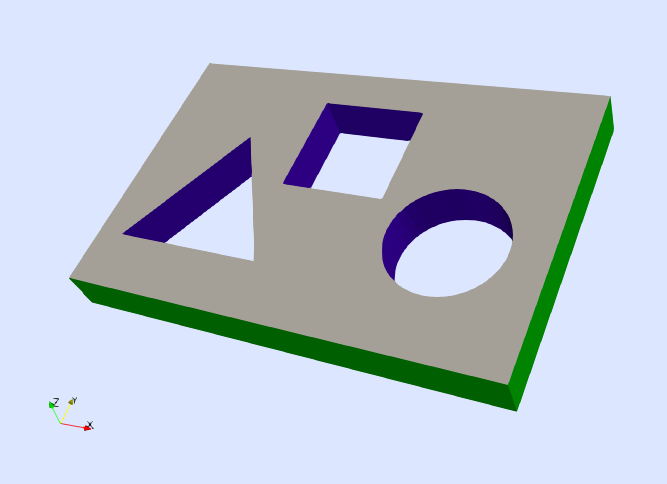}
\caption{\tutorial: View of the computational domain.}
\label{fig:p_laplacian_model}
\end{figure}

The goal of this tutorial is to solve a nonlinear \ac{pde} in \gridap. For the sake of simplicity, we consider the $p$-Laplacian equation as the model problem. Specifically, the \ac{pde}  we want to solve is: find the scalar-field $u$ such that
\begin{align}
\left\lbrace
\begin{aligned}
-\nabla \cdot \left( |\nabla u|^{p-2} \ \nabla u \right) = f\ &\text{in}\ \Omega,\\
u = 0 \ &\text{on} \ \Gamma_0,\\
u = g \ &\text{on} \ \Gamma_g,\\
\left( |\nabla u|^{p-2}\ \nabla u \right)\cdot n = 0 \ &\text{on} \ \Gamma_{\rm N},
\end{aligned}
\right.
\end{align}
with $p>2$.
The computational domain $\Omega$ is the one depicted in \fig{fig:p_laplacian_model}, which is the same as in the first tutorial. However, we slightly change the boundary conditions here. We impose homogeneous Dirichlet and homogeneous Neumann boundary conditions on $\Gamma_0$ and $\Gamma_{\rm N}$  respectively, and in-homogeneous Dirichlet conditions on $\Gamma_g$. The Dirichlet boundaries $\Gamma_0$ and $\Gamma_g$ are defined as the closure of the green and blue surfaces in \fig{fig:p_laplacian_model} respectively, whereas the Neumann boundary is $\Gamma_{\rm N}\doteq\partial\Omega \setminus (\Gamma_0\cup\Gamma_g)$. In this example, we consider the values $p=3$, $f=1$, and $g=2$.

\subsection{Numerical scheme}

We discretize the problem with conforming Lagrangian FE spaces. For this formulation, the nonlinear weak form reads: find $u\in U_g$ such that $[r(u)](v) = 0$ for all $v\in V_0$. As in previous tutorials, the space $U_g$ is the set of functions in $H^1(\Omega)$ that fulfill the Dirichlet boundary conditions, whereas $V_0$ is composed by functions in $H^1(\Omega)$ that vanish at the Dirichlet boundary. The weak residual $r(u)$ evaluated at a function  $u\in U_g$ is the linear form defined as
\begin{equation}
[r(u)](v) \doteq \int_\Omega \nabla v \cdot \left( |\nabla u|^{p-2}\ \nabla u \right) \ {\rm d}\Omega - \int_\Omega v\ f \ {\rm d}\Omega.
\end{equation}

In order to solve this nonlinear weak equation, we consider a Newton-Raphson method, which is associated with a linearization of the problem in an arbitrary direction $\delta u\in V_0$, namely $[r(u+\delta u)](v)\approx [r(u)](v) + [j(u)](v,\delta u)$. In previous formula,  $j(u)$ is the Jacobian evaluated at $u\in U_g$, which is the bilinear form
\begin{equation}
[j(u)](\delta u,v) = \int_\Omega \nabla v \cdot \left( |\nabla u|^{p-2}\ \nabla \delta u \right) \ {\rm d}\Omega + (p-2) \int_\Omega \nabla v \cdot \left(  |\nabla u|^{p-4} (\nabla u \cdot \nabla \delta u) \nabla u  \right) \ {\rm d}\Omega.
\end{equation}

Note that the solution of this nonlinear \ac{pde} with a Newton-Raphson method, will require to discretize both the residual $r$ and the Jacobian $j$. In \gridap, this is done by following an approach similar to the one already shown in previous tutorials for discretizing the bilinear and linear forms associated with a linear FE problem. The specific details are discussed below.

\subsection{Discrete model}

First, we load a discretization of the computational domain. In this case, we the same discrete model as in the first tutorial

\begin{minted}[bgcolor=bg]{julia1}
using Gridap
model = DiscreteModelFromFile("model.json")
\end{minted}

%

\begin{figure}[ht!]
\includegraphics[width=0.6\textwidth]{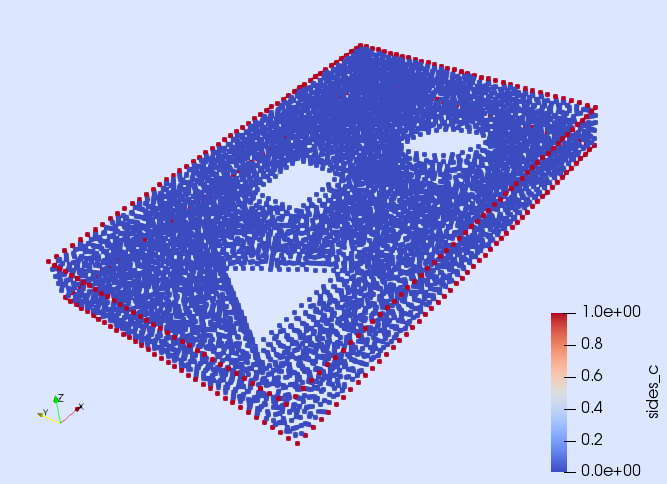}
\caption{\tutorial: View of the vertices of the discrete model. Vertices on the contour of the Dirichlet boundary $\Gamma_0$ are identified with the \shb{"sides\_c"} tag.}
\label{fig:p_laplacian_sides_c}
\end{figure}

As stated before, we want to impose Dirichlet boundary conditions on $\Gamma_0$ and $\Gamma_g$,  but none of these boundaries is identified in the model. E.g., you can easily see by writing the model in vtk format 
\begin{minted}[bgcolor=bg]{julia1}
writevtk(model,"model")
\end{minted}
and by opening the file \shb{"model\_0"} in \paraview{} that the boundary identified as \shb{"sides"} only includes the vertices in the interior of $\Gamma_0$, but here we want to impose Dirichlet boundary conditions in the closure of $\Gamma_0$, i.e., also on the vertices on the contour of $\Gamma_0$. Fortunately, the objects on the contour of $\Gamma_0$ are identified  with the tag \shb{"sides\_c"} (see \fig{fig:p_laplacian_sides_c}). Thus, the Dirichlet boundary $\Gamma_0$ can be built as the union of the objects identified as \shb{"sides"} and \shb{"sides\_c"}.

\gridap{} provides a convenient way to create new object identifiers (referred to as ``tags") from existing ones. First, we need to extract from the model, the object that holds the information about the boundary identifiers (referred to as ``face labeling"):
\begin{minted}[bgcolor=bg]{julia1}
labels = get_face_labeling(model)
\end{minted}
Then, we can add new identifiers (aka "tags") to it. In the next line, we create a new tag called \shb{"diri0"} as the union of the objects identified as \shb{"sides"} and \shb{"sides\_c"}, which is precisely what we need to represent the closure of the Dirichlet boundary $\Gamma_0$.

\begin{minted}[bgcolor=bg]{julia1}
add_tag_from_tags!(labels,"diri0",["sides", "sides_c"])
\end{minted}

We follow the same approach to build a new identifier for the closure of the Dirichlet boundary $\Gamma_g$. In this case, the boundary is expressed as the union of the objects identified with the tags \shb{"circle"}, \shb{"circle\_c"}, \shb{"triangle"}, \shb{"triangle\_c"}, \shb{"square"}, \shb{"square\_c"}. Thus, we create a new tag for  $\Gamma_g$, called \shb{"dirig"} simply as follows:

\begin{minted}[bgcolor=bg]{julia1}
add_tag_from_tags!(labels,"dirig",
  ["circle","circle_c", "triangle", "triangle_c", "square", "square_c"])
\end{minted}

\subsection{FE Space}

Now, we can build the FE space by using the newly defined boundary tags.

\begin{minted}[bgcolor=bg]{julia1}
V0 = TestFESpace(
  reffe=:Lagrangian, order=1, valuetype=Float64,
  conformity=:H1, model=model, labels=labels,
  dirichlet_tags=["diri0", "dirig"])
\end{minted}
The construction of this space is essentially the same as in the first tutorial (we build a continuous scalar-valued Lagrangian interpolation of first order). However, we also pass here the \shb{labels} object (that contains the newly created boundary tags). From this FE space, we define the trial FE spaces

\begin{minted}[bgcolor=bg]{julia1}
g = 1
Ug = TrialFESpace(V0,[0,g])
\end{minted}

\subsection{Nonlinear FE problem}

At this point, we are ready to build the nonlinear FE problem. To this end, we need to define the weak residual and also its corresponding Jacobian. This is done following a similar procedure to the one considered in previous tutorials to define the bilinear and linear forms associated with linear FE problems. In this case, instead of an \shb{AffineFETerm} (which is for linear problems), we use a \shb{FETerm}. An instance of \shb{FETerm} is constructed by providing the integrands of the weak residual and its Jacobian (in a similar way an \shb{AffineFETerm} is constructed from the integrands of the bilinear and linear forms). 

On the one hand, the integrand of the weak residual is built as follows

\begin{minted}[bgcolor=bg]{julia1}
using LinearAlgebra: norm
const p = 3
@law flux(∇u) = norm(∇u)^(p-2) * ∇u
f(x) = 1
res(u,v) = ∇(v)*flux(∇(u)) - v*f
\end{minted}
Function \shb{res} is the one representing the integrand of the weak residual $[r(u)](v)$. The first argument of function \shb{res} stands for the function $u\in U_g$, where the residual is evaluated, and the second argument stands for a generic test function $v\in V_0$. Note that we have used the macro \shb{@law} to construct the ``constitutive  law" that relates the nonlinear flux with the gradient of the solution.

On the other hand,  we implement a function \shb{jac} representing the integrand of the Jacobian
\begin{minted}[bgcolor=bg]{julia1}
@law dflux(∇du,∇u) =
  (p-2)*norm(∇u)^(p-4)*inner(∇u,∇du)*∇u + norm(∇u)^(p-2) * ∇du
jac(u,du,v) = ∇(v)*dflux(∇(du),∇(u))
\end{minted}
The first argument of function \shb{jac} stands for function $u\in U_g$, where the Jacobian is evaluated. The second argument represents an arbitrary direction $\delta u \in V_0$, and the third argument is a generic test function $v\in V_0$. Note that we have also used the macro \shb{@law} to define the linearization of the nonlinear flux. 

With these functions, we build the \shb{FETerm} as follows:
\begin{minted}[bgcolor=bg]{julia1}
trian = Triangulation(model)
degree=2
quad = CellQuadrature(trian,degree)
t_Ω = FETerm(res,jac,trian,quad)
\end{minted}
We build the \shb{FETerm} by passing in the first and second arguments the functions that represent the integrands of the residual and Jacobian respectively. The other two arguments, are the triangulation and quadrature used to perform the integrals numerically. From this \shb{FETerm} object, we finally construct the nonlinear FE problem

\begin{minted}[bgcolor=bg]{julia1}
op = FEOperator(Ug,V0,t_Ω)
\end{minted}
Here, we have constructed an instance of \shb{FEOperator}, which is the type that represents a general nonlinear FE problem in \gridap. The constructor takes the test and trial spaces, and the \shb{FETerms} objects describing the corresponding weak form (in this case only a single term).

\subsection{Nonlinear solver phase}

We have already built the nonlinear FE problem. Now, the remaining step is to solve it. In \gridap, nonlinear (and also linear) FE problems can be solved with instances of the type \shb{FESolver}. The type \shb{FESolver} is a concrete implementation of the abstract type \shb{FESolver} particularly designed for nonlinear problems (in contrast to the concrete type \shb{LinearFESolver} which is for the linear case).
%
%
%
We construct an instance of \shb{FESolver} as follows:
\begin{minted}[bgcolor=bg]{julia1}
using LineSearches: BackTracking
nls = NLSolver(
  show_trace=true, method=:newton, linesearch=BackTracking())
solver = FESolver(nls)
\end{minted}
Note that the \shb{NLSolver} function used above internally calls the \shb{nlsolve} function of the {NLsolve} package \cite{Carlsson_NLsolve_2019} with the provided key-word arguments. Thus, one can use any of the nonlinear methods available via the function \shb{nlsolve} to solve the nonlinear \ac{fe} problem. Here, we have selected a Newton-Raphson method with a back-tracking line-search from the {LineSearches} package \cite{linesearches_web}.



\begin{figure}[ht!]
\includegraphics[width=0.6\textwidth]{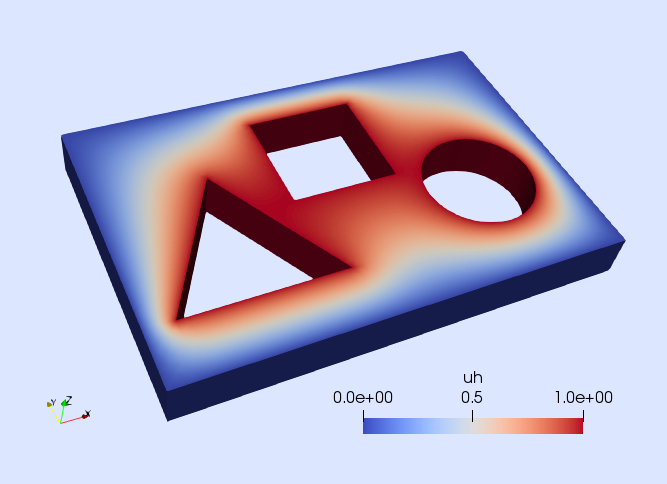}
\caption{\tutorial: View of the computed numerical solution.}
\label{fig:p_laplacian_sol-plap}
\end{figure}

We are finally in place to solve the nonlinear \ac{fe} problem. The initial guess is a \shb{FEFunction}, which we build from a vector of random (free) nodal values:
\begin{minted}[bgcolor=bg]{julia1}
import Random
Random.seed!(1234)
x = rand(Float64,num_free_dofs(Ug))
uh0 = FEFunction(Ug,x)
uh, = solve!(uh0,solver,op)
\end{minted}
%
%
%
%

We finish this tutorial by writing the computed solution for visualization (see \fig{fig:p_laplacian_sol-plap}).    
\begin{minted}[bgcolor=bg]{julia1}
writevtk(trian,"results",cellfields=["uh"=>uh])
\end{minted}

\stepcounter{tutorial}
\section{Discontinuous Galerkin Methods (\tutorial)}
In this tutorial, we will learn
\begin{itemize}
  \item How to solve a simple \ac{pde} with a \ac{dg} method
  \item How to compute jumps and averages of quantities on the mesh skeleton
  \item How to implement the method of manufactured solutions
  \item How to integrate error norms
  \item How to generate Cartesian meshes in arbitrary dimensions
\end{itemize}

\subsection{Problem statement}

The goal of this tutorial is to solve a \ac{pde} using a \ac{dg} formulation. For simplicity, we take the Poisson equation on the unit cube $\Omega \doteq (0,1)^3$ as the model problem, namely
\begin{align}
\left\lbrace
\begin{aligned}
-\Delta u = f  \ &\text{in} \ \Omega,\\
u = g \ &\text{on}\ \partial\Omega,\\
\end{aligned}
\right.
\end{align}
where $f$ is the source term and $g$ is the prescribed Dirichlet boundary function. In this tutorial, we follow the method of manufactured solutions since we want to illustrate how to compute discretization errors. We take $u(x) = 3 x_1 + x_2 + 2 x_3$ as the exact solution of the problem, for which $f=0$ and $g(x) = u(x)$ are computed. The selected manufactured solution $u$ is a first order multi-variate polynomial, which can be represented exactly by the \ac{fe} interpolation that we are going to define below. In this scenario, the discretization error has to be close to the machine precision. We will use this result to validate the proposed implementation.



\subsection{Numerical Scheme}

We consider a \ac{dg} formulation to approximate
the problem. In particular, we consider  the symmetric
interior penalty method (see, e.g. \cite{Arnold2001}, for specific details). For this formulation, the approximation space is made of discontinuous piece-wise polynomials, namely
\begin{equation}
V \doteq \{ v\in L^2(\Omega):\ v|_{T}\in Q_p(T) \text{ for all } T\in\mathcal{T}  \},
\end{equation}
where $\mathcal{T}$ is the set of all cells $T$ of the FE mesh, and $Q_p(T)$ is a polynomial space of degree $p$ defined on a generic cell $T$. For simplicity, we consider Cartesian meshes in this tutorial. In this case, the space $Q_p(T)$ is made of multi-variate polynomials up to degree $p$ in each spatial coordinate.

In order to write the weak form of the problem, we need to introduce some notation. The sets of interior and boundary facets associated with the \ac{fe} mesh $\mathcal{T}$ are denoted here as $\mathcal{F}_\Gamma$ and $\mathcal{F}_{\partial\Omega}$ respectively. In addition, for a given function $v\in V$ restricted to the interior facets $\mathcal{F}_\Gamma$, we introduce the well known jump and mean value operators,
\begin{equation}
\lbrack\!\lbrack v\ n \rbrack\!\rbrack \doteq v^+\ n^+ + v^- n^-, \text{ and } \{\! \!\{\nabla v\}\! \!\} \doteq \dfrac{ \nabla v^+ + \nabla v^-}{2},
\end{equation}
with $v^+$, and $v^-$ being the restrictions of $v\in V$ to the cells $T^+$, $T^-$ that share a generic interior facet in $\mathcal{F}_\Gamma$, and $n^+$, and $n^-$ are the facet outward unit normals from either the perspective of $T^+$ and $T^-$ respectively.

With this notation, the weak form associated with the interior penalty formulation of our problem reads: find $u\in V$ such that $a(u,v) = b(v)$ for all $v\in V$.  The bilinear and linear forms  $a(\cdot,\cdot)$ and $b(\cdot)$ have contributions associated with the bulk of $\Omega$,  the boundary facets $\mathcal{F}_{\partial\Omega}$, and the interior facets  $\mathcal{F}_\Gamma$, namely
 \begin{align}
\begin{aligned}
a(u,v) &= a_{\Omega}(u,v) + a_{\partial\Omega}(u,v) + a_{\Gamma}(u,v),\\
b(v) &= b_{\Omega}(v) + b_{\partial\Omega}(v).
\end{aligned}
\end{align}
These contributions are defined for the volume,
\begin{align}
a_{\Omega}(u,v) \doteq \sum_{T\in\mathcal{T}} \int_{T} \nabla v \cdot \nabla u \ {\rm d}T, \quad b_{\Omega}(v) \doteq \int_{\Omega} v\ f \ {\rm d}\Omega,
 \end{align}
for the boundary facets,
\begin{align}
\begin{aligned}
a_{\partial\Omega}(u,v) &\doteq \sum_{F\in\mathcal{F}_{\partial\Omega}} \dfrac{\gamma}{|F|} \int_{F} v\ u \ {\rm d}F -  \sum_{F\in\mathcal{F}_{\partial\Omega}} \int_{F} v\ (\nabla u \cdot n)  \ {\rm d}F -  \sum_{F\in\mathcal{F}_{\partial\Omega}} \int_{F} (\nabla v \cdot n)\ u  \ {\rm d}F, \\
b_{\partial\Omega} &\doteq \sum_{F\in\mathcal{F}_{\partial\Omega}} \dfrac{\gamma}{|F|} \int_{F} v\ g \ {\rm d}F  -  \sum_{F\in\mathcal{F}_{\partial\Omega}} \int_{F} (\nabla v \cdot n)\ g  \ {\rm d}F,
\end{aligned}
 \end{align}
and for the interior facets,
\begin{align}
a_{\Gamma}(u,v) \doteq \sum_{F\in\mathcal{F}_{\Gamma}} \dfrac{\gamma}{|F|} \int_{F} \lbrack\!\lbrack v\ n \rbrack\!\rbrack\cdot \lbrack\!\lbrack u\ n \rbrack\!\rbrack \ {\rm d}F -  \sum_{F\in\mathcal{F}_{\Gamma}} \int_{F} \lbrack\!\lbrack v\ n \rbrack\!\rbrack\cdot \{\! \!\{\nabla u\}\! \!\} \ {\rm d}F -  \sum_{F\in\mathcal{F}_{\Gamma}} \int_{F} \{\! \!\{\nabla v\}\! \!\}\cdot \lbrack\!\lbrack u\ n \rbrack\!\rbrack \ {\rm d}F.
 \end{align}
 In previous expressions, $|F|$ denotes the diameter of the face $F$ (in our Cartesian grid, this is equivalent to the characteristic mesh size $h$), and $\gamma$ is a stabilization parameter that should be chosen large enough such that the bilinear form $a(\cdot,\cdot)$ is stable and continuous. Here, we take $\gamma = p\ (p+1)$ as done in the numerical experiments in reference \cite{Cockburn2009}.

\subsection{Manufactured solution}


We start by loading the \gridap{} library and defining the manufactured solution $u$ and the associated source term $f$ and Dirichlet function $g$.
\begin{minted}[bgcolor=bg]{julia1}
using Gridap
u(x) = 3*x[1] + x[2] + 2*x[3]
f(x) = 0
g(x) = u(x)
\end{minted}
We also need to define the gradient of $u$ since we will compute the $H^1$ error norm later. In that case, the gradient is simply defined as
\begin{minted}[bgcolor=bg]{julia1}
∇u(x) = VectorValue(3,1,2)
\end{minted}
In addition, we need to tell the \gridap{} library that the gradient of the function \shb{u} is available in the function \shb{$\nabla$u} (at this moment \shb{u} and \shb{$\nabla$u} are two standard Julia functions without any connection between them). This is done by adding an extra method to the function \shb{gradient} (aka \shb{$\nabla$}) defined in \gridap:
\begin{minted}[bgcolor=bg]{julia1}
import Gridap: ∇
∇(::typeof(u)) = ∇u
\end{minted}
 Now, it is possible to recover function \shb{$\nabla$u} from function \shb{u} as \shb{$\nabla$(u)}. You can check that the following expression evaluates to \shb{true}.
\begin{minted}[bgcolor=bg]{julia1}
∇(u) === ∇u
\end{minted}

\subsection{Cartesian mesh generation}
 In order to discretize the geometry of the unit cube, we use the Cartesian mesh generator available in \gridap{}.

\begin{minted}[bgcolor=bg]{julia1}
L = 1.0
domain = (0.0, L, 0.0, L, 0.0, L)
n = 4
partition = (n,n,n)
model = CartesianDiscreteModel(domain,partition)
\end{minted}

\begin{figure}[ht!]
\includegraphics[width=0.6\textwidth]{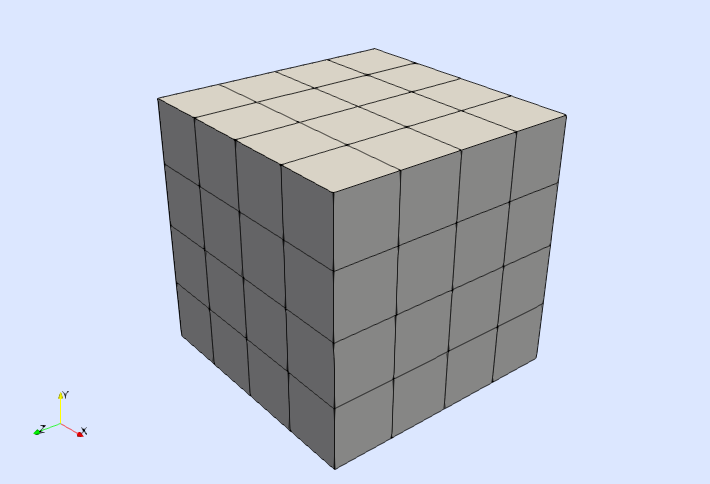}
\caption{\tutorial: View of the Cartesian discrete model.}
\label{fig:poisson_dg_model}
\end{figure}

The type \shb{CartesianDiscreteModel} is a concrete type that inherits from \shb{DiscreteModel}, which is specifically designed for building Cartesian meshes. The \shb{CartesianDiscreteModel} constructor takes a tuple containing limits of the box we want to discretize  plus a tuple with the number of cells to be generated in each direction (here $4\times4\times4$ cells). You can write the model in vtk format to visualize it (see \fig{fig:poisson_dg_model}). 
\begin{minted}[bgcolor=bg]{julia1}
writevtk(model,"model")
\end{minted}

 Note that the \shb{CaresianDiscreteModel} is implemented for arbitrary dimensions. For instance, the following lines build a \shb{CartesianDiscreteModel}  for the unit square $(0,1)^2$ with 4 cells per direction
\begin{minted}[bgcolor=bg]{julia1}
domain2D = (0.0, L, 0.0, L)
partition2D = (n,n)
model2D = CartesianDiscreteModel(domain2D,partition2D)
\end{minted}
You could also generate a mesh for the unit tesseract $(0,1)^4$ (i.e., the unit cube in 4D) by extrapolating how 2D and 3D models are constructed.

\subsection{\ac{fe} spaces} 

On top of the discrete model, we create the discontinuous space $V$ as follows
\begin{minted}[bgcolor=bg]{julia1}
order = 3
V = TestFESpace(
  reffe=:Lagrangian, valuetype=Float64, order=order,
  conformity=:L2, model=model)
\end{minted}
We have select a Lagrangian, scalar-valued interpolation of order $3$ within the cells of the discrete model. Since the cells are hexahedra, the resulting Lagrangian shape functions are tri-cubic polynomials. In contrast to previous tutorials, where we have constructed $H^1$-conforming (i.e., continuous) \ac{fe} spaces, here we construct a $L^2$-conforming (i.e., discontinuous) \ac{fe} space. That is, we do not impose any type of continuity of the shape function on the cell boundaries, which leads to the discontinuous \ac{fe} space $V$ of the \ac{dg} formulation. Note also that we do not pass any information about the Dirichlet boundary to the \shb{TestFESpace} constructor since the Dirichlet boundary conditions are not imposed strongly in this example.

From the \shb{V} object we have constructed in previous code snippet, we build the trial \ac{fe} space.
\begin{minted}[bgcolor=bg]{julia1}
U = TrialFESpace(V)
\end{minted}
Note that we do not pass any Dirichlet function to the \shb{TrialFESpace} constructor since we do not impose Dirichlet boundary conditions strongly in this example.

\subsection{Numerical integration} Once the \ac{fe} spaces are ready, the next step is to set up  the numerical integration. In this example, we need to integrate in three different domains: the volume covered by the cells $\mathcal{T}$  (i.e., the computational domain $\Omega$), the surface covered by the boundary facets $\mathcal{F}_{\partial\Omega}$ (i.e., the boundary $\partial\Omega$), and the surface covered by the interior facets $\mathcal{F}_{\Gamma}$ (i.e. the so-called mesh skeleton). In order to integrate in $\Omega$ and on its boundary $\partial\Omega$, we use \shb{Triangulation} and \shb{BoundaryTriangulation} objects as already discussed in previous tutorials.

\begin{minted}[bgcolor=bg]{julia1}
trian = Triangulation(model)
btrian = BoundaryTriangulation(model)
\end{minted}
Note that we have not passed any boundary identifier to the \shb{BoundaryTriangulation} constructor. In this case, an integration mesh for the entire boundary $\partial\Omega$ is constructed by default (which is what we need in this example).

\begin{figure}[ht!]
\includegraphics[width=0.6\textwidth]{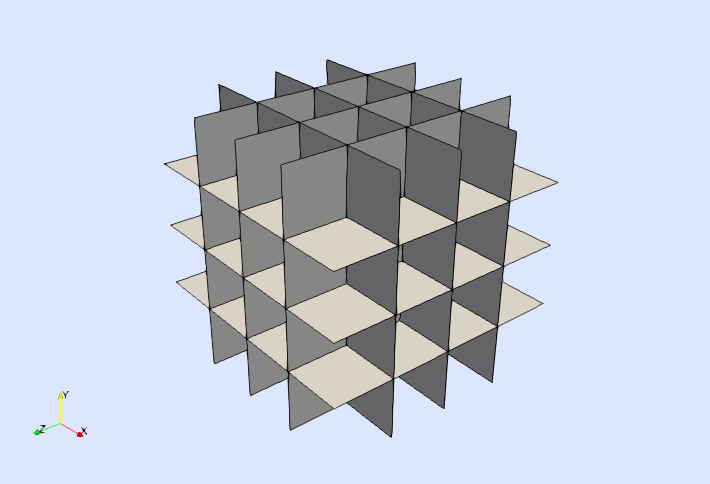}
\caption{\tutorial: View of the interior facets.}
\label{fig:poisson_dg_skeleton_trian}
\end{figure}

In order to generate an integration mesh for the interior facets $\mathcal{F}_{\Gamma}$, we use a new type of \shb{Triangulation} referred to as \shb{SkeletonTriangulation}. It can be constructed from a\linebreak \shb{DiscreteModel} object as follows:
\begin{minted}[bgcolor=bg]{julia1}
strian = SkeletonTriangulation(model)
\end{minted}
As any other type of \shb{Triangulation}, an \shb{SkeletonTriangulation} can be written into a vtk file for its visualization (see \fig{fig:poisson_dg_skeleton_trian}, where the interior facets $\mathcal{F}_\Gamma$ are clearly observed).
\begin{minted}[bgcolor=bg]{julia1}
writevtk(strian,"strian")
\end{minted}

Once we have constructed the triangulations needed in this example, we define the corresponding quadrature rules.
\begin{minted}[bgcolor=bg]{julia1}
degree = 2*order
quad = CellQuadrature(trian,degree)
bquad = CellQuadrature(btrian,degree)
squad = CellQuadrature(strian,degree)
\end{minted}

We still need a way to represent the unit outward normal vector to the boundary $\partial\Omega$, and the unit normal vector on the interior faces $\mathcal{F}_\Gamma$. This is done with the \shb{get\_normal\_vector} function.
\begin{minted}[bgcolor=bg]{julia1}
nb = get_normal_vector(btrian)
ns = get_normal_vector(strian)
\end{minted}
The \shb{get\_normal\_vector} function takes either a boundary or a skeleton triangulation and returns an object representing the normal vector to the corresponding surface. For boundary triangulations, the returned normal vector is the unit outwards one, whereas for skeleton triangulations the orientation of the returned normal is arbitrary. In the current implementation (\gridap{} v0.8.0), the unit normal is outwards to the cell with smaller id among the two cells that share an interior facet in $\mathcal{F}_\Gamma$.

\subsection{Weak form} With these ingredients we can define the different terms in the weak form. First, we start with the terms $a_\Omega(\cdot,\cdot)$ , and $b_\Omega(\cdot)$ associated with integrals in the volume $\Omega$. This is done as in the tutorial for the Poisson equation.

\begin{minted}[bgcolor=bg]{julia1}
a_Ω(u,v) = ∇(v)*∇(u)
b_Ω(v) = v*f
t_Ω = AffineFETerm(a_Ω,b_Ω,trian,quad)
\end{minted}

The terms $a_{\partial\Omega}(\cdot,\cdot)$ and $b_{\partial\Omega}(\cdot)$ associated with integrals on the boundary $\partial\Omega$ are defined using an analogous approach. First, we define two functions representing the integrands of the forms $a_{\partial\Omega}(\cdot,\cdot)$ and $b_{\partial\Omega}(\cdot)$. Then, we build an \shb{AffineFETerm} from these functions and the boundary triangulation and its corresponding quadrature rule:

\begin{minted}[bgcolor=bg]{julia1}
h = L / n
γ = order*(order+1)
a_∂Ω(u,v) = (γ/h)*v*u - v*(∇(u)*nb) - (∇(v)*nb)*u
b_∂Ω(v) = (γ/h)*v*g - (∇(v)*nb)*g
t_∂Ω = AffineFETerm(a_∂Ω,b_∂Ω,btrian,bquad)
\end{minted}
Note that in the definition of the functions \shb{a\_∂Ω} and \shb{b\_∂Ω}, we have used the object \shb{nb} representing the outward unit normal to the boundary $\partial\Omega$. The code definition of  \shb{a\_∂Ω} and \shb{b\_∂Ω} is indeed very close to the mathematical definition of the forms  $a_{\partial\Omega}(\cdot,\cdot)$ and $b_{\partial\Omega}(\cdot)$. 

Finally, we need to define the term $a_\Gamma(\cdot,\cdot)$ integrated on the interior facets $\mathcal{F}_\Gamma$. In this case, we use a \shb{LinearFETerm} since the terms integrated on the interior facets only contribute to the system matrix and not to the right-hand-side vector.
\begin{minted}[bgcolor=bg]{julia1}
a_Γ(u,v) = (γ/h)*jump(v*ns)*jump(u*ns) -
   jump(v*ns)*mean(∇(u)) - mean(∇(v))*jump(u*ns)
t_Γ = LinearFETerm(a_Γ,strian,squad)
\end{minted} 
Note that the arguments \shb{u}, \shb{v} of function  \shb{a\_Γ} represent a trial and test function \emph{restricted} to the interior facets $\mathcal{F}_\Gamma$. As mentioned before in the presentation of the \ac{dg} formulation, the restriction of a function $v\in V$ to the interior faces leads to two different values $v^+$ and $v^-$ . In order to compute jumps and averages of the quantities $v^+$ and $v^-$, we use the functions \shb{jump} and \shb{mean}, which represent the jump and mean value operators $\lbrack\!\lbrack \cdot \rbrack\!\rbrack$ and $\{\! \!\{\cdot\}\! \!\}$ respectively. Note also that we have used the object \shb{ns} representing the unit normal vector on the interior facets. As a result, the notation used to define function \shb{a\_Γ} is very close to the mathematical definition of the terms in the bilinear form $a_\Gamma(\cdot,\cdot)$. 

Once the different terms of the weak form have been defined, we build and solve the FE problem.
\begin{minted}[bgcolor=bg]{julia1}  
op = AffineFEOperator(U,V,t_Ω,t_∂Ω,t_Γ)
uh = solve(op)
\end{minted}

\subsection{Discretization error} 

\begin{figure}[ht!]
\includegraphics[width=0.6\textwidth]{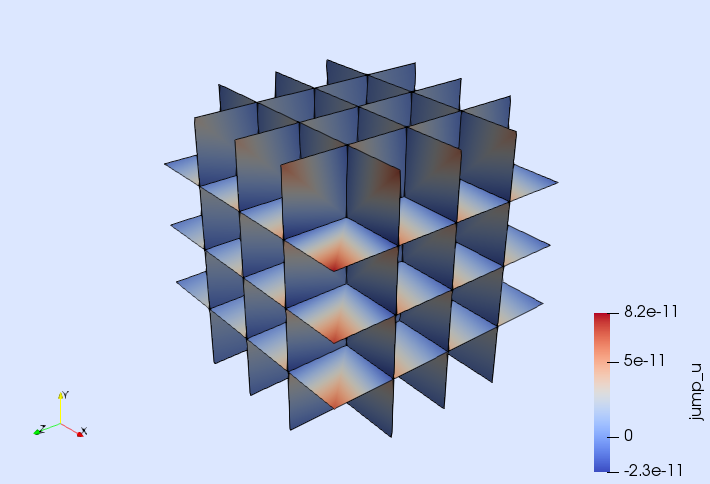}
\caption{\tutorial: View of the jump of the numerical solution at the interior facets.}
\label{fig:poisson_dg_jump_u}
\end{figure}

We end this tutorial by quantifying the discretization error associated with  the computed numerical solution \shb{uh}. In \ac{dg} methods, a simple error indicator is the jump of the computed (discontinuous) approximation on the interior faces. This quantity can be easily computed in \gridap{} as follows. First, we need to restrict the computed solution \shb{uh} to the skeleton triangulation.
\begin{minted}[bgcolor=bg]{julia1} 
uh_Γ = restrict(uh,strian)
\end{minted}
The resulting object \shb{uh\_Γ} is an object which represents the two values $u^+_h$, $u^-_h$ of the solution $u_h$ restricted to the interior facets $\mathcal{F}_\Gamma$. We compute and visualize the jump of these values as follows (see \fig{fig:poisson_dg_jump_u}):
\begin{minted}[bgcolor=bg]{julia1} 
writevtk(strian,"jumps",cellfields=["jump_u"=>jump(uh_Γ)])
\end{minted}
Note that the jump of the numerical solution is very small, close to the machine precision (as expected in this example with manufactured solution).

 A more rigorous way of quantifying the error is to measure it with a norm. Here, we use the $L^2$ and $H^1$ norms, namely
 \begin{equation}
 \| w \|_{L^2}^2 \doteq \int_{\Omega} w^2 \ \text{d}\Omega, \quad
 \| w \|_{H^1}^2 \doteq \int_{\Omega} w^2 + \nabla w \cdot \nabla w \ \text{d}\Omega.
\end{equation}

The discretization error can be computed in this example as the difference of the manufactured and numerical solutions.
\begin{minted}[bgcolor=bg]{julia1}
e = u - uh
\end{minted}
We compute the error norms as follows. First, we implement the integrands of the norms we want to compute.
\begin{minted}[bgcolor=bg]{julia1}
l2(u) = inner(u,u)
h1(u) = a_Ω(u,u) + l2(u)
\end{minted}
Then, we compute the corresponding integrals with the \shb{integrate} function.
\begin{minted}[bgcolor=bg]{julia1}
el2 = sqrt(sum( integrate(l2(e),trian,quad) ))
eh1 = sqrt(sum( integrate(h1(e),trian,quad) ))
\end{minted}
The \shb{integrate} function returns a lazy object representing the contribution to the integral of each cell in the underlying triangulation. To end up with the desired error norms, one has to sum these contributions and take the square root. You can check that the computed error norms are close to machine precision (as one would expect).

\begin{minted}[bgcolor=bg]{julia1}
tol = 1.e-10
@assert el2 < tol
@assert eh1 < tol
\end{minted}

\stepcounter{tutorial}
\section{Darcy problem (\tutorial)} \label{sec:tutorial03}


In this tutorial, we will learn
\begin{itemize}
   \item How to implement multi-field \acp{pde}
   \item How to build div-conforming \ac{fe} spaces
   \item How to impose boundary conditions in multi-field problems
\end{itemize}

\subsection{Problem statement}

\def\u{\boldsymbol{u}}
\def\x{\boldsymbol{x}}
\def\v{\boldsymbol{v}}
\def\n{\boldsymbol{n}}
\def\grad{\boldsymbol{\nabla}}

In this tutorial, we show how to solve a multi-field \ac{pde} in \gridap{}. As a model problem, we consider the Darcy equations with Dirichlet and Neumann boundary conditions. The \ac{pde} to solve is: find the flux vector $u$, and the pressure $p$ such that 
\begin{align}
   \left\lbrace
   \begin{aligned}
      \kappa^{-1} u + \nabla p = 0  \ &\text{in} \ \Omega,\\
      \nabla \cdot u = f  \ &\text{in} \ \Omega,\\
      u \cdot n = g \ &\text{on}\  \Gamma_{\rm D},\\
      p = h \ &\text{on}\ \Gamma_{\rm N},\\
   \end{aligned}
   \right.
\end{align}
being $n$ the outwards unit normal vector to the boundary $\partial\Omega$.  We consider the unit square $\Omega \doteq (0,1)^2$ as the computational domain, the Neumann boundary $\Gamma_{\rm N}$ is the right and left sides of $\Omega$, and $\Gamma_{\rm D}$ is the bottom and top sides of $\Omega$. We consider $f = g \doteq 0$ and $h(x) \doteq x_1$, i.e., $h$ equal to 0 on the left side and 1 on the right side. The inverse of the permeability tensor, namely $\kappa^{-1}(x)$, is chosen equal to
$$
\left( \begin{tabular}{ll}
  100 & 90 \\
  90 & 100
\end{tabular} \right)  \text{ for } \ x \in [0.4,0.6]^2, \text{ and }
\left( \begin{tabular}{ll}
  1 & 0 \\
  0 & 1
\end{tabular} \right) \ \text	{otherwise.}
$$

In order to state this problem in weak form, we introduce the following Sobolev spaces. $H(\mathrm{div};\Omega)$ is the space of vector fields in $\Omega$, whose components and divergence are in $L^2(\Omega)$. On the other hand, $H_g(\mathrm{div};\Omega)$ and $H_0(\mathrm{div};\Omega)$ are the subspaces of functions in $H(\mathrm{div};\Omega)$ such that their normal traces are equal to $g$ and $0$ respectively almost everywhere in $\Gamma_{\rm D}$. With these notations, the weak form reads: find $(u,p)\in H_g(\mathrm{div};\Omega)\times L^2(\Omega)$ such that $a((u,q),(v,q)) = b(v,q)$ for all $(v,q)\in H_0(\mathrm{div};\Omega)\times L^2(\Omega)$, where
\begin{align}
\begin{aligned}
a((u,p),(v,q)) &\doteq \int_{\Omega}  v \cdot \left(\kappa^{-1} u\right) \ {\rm d}\Omega - \int_{\Omega} (\nabla \cdot v)\ p \ {\rm d}\Omega + \int_{\Omega} q\ (\nabla \cdot u) \ {\rm d}\Omega,\\
b(v,q) &\doteq \int_{\Omega} q\ f \ {\rm  d}\Omega - \int_{\Gamma_{\rm N}} (v\cdot n)\ h  \ {\rm  d}\Gamma.
\end{aligned}
\end{align}

 \subsection{Numerical scheme}

In this tutorial, we use the \ac{rt}  space for the flux approximation \cite{brezzi_mixed_1991}. On a reference square with sides aligned with the Cartesian axes, the \ac{rt} space of order $k$ is represented as $Q_{(k+1,k)} \times Q_{(k,k+1)}$, being the polynomial space defined as follows. The component  $w_\alpha$ of a vector field $w$ in $Q_{(k+1,k)} \times Q_{(k,k+1)}$ is obtained as the tensor product of univariate polynomials of order $k+1$ in direction $\alpha$ times univariate polynomials of order $k$ on the other directions. That is, $\nabla\cdot w \in Q_k$, where $Q_k$ is the multivariate polynomial space of degree at most $k$ in each of the spatial coordinates. Note that the definition of the \ac{rt} space also applies to arbitrary dimensions. The global \ac{fe} space for the flux $V$ is obtained by mapping the cell-wise \ac{rt} space into the physical space using the Piola transformation and enforcing continuity of normal traces across cells (see \cite{brezzi_mixed_1991} for specific details). 
 We consider the subspace  $V_0$ of functions in $V$ with zero normal trace on $\Gamma_{\rm D}$, and the subspace $V_g$ of functions in $V$ with normal trace equal to the projection of $g$ onto the space of traces of $V$ on $\Gamma_{\rm D}$. With regard to the pressure, we consider the discontinuous space of cell-wise polynomials in $Q_k$.

\subsection{Discrete model}

We start the driver loading the \gridap{} package and constructing the geometrical model. We generate a $100\times100$ structured mesh for the domain $(0,1)^2$.

\begin{minted}[bgcolor=bg]{julia1}
using Gridap
domain = (0,1,0,1)
partition = (100,100)
model = CartesianDiscreteModel(domain,partition)
\end{minted}

\subsection{Multi-field \ac{fe} spaces}

Next, we build the \ac{fe} spaces. We consider the first order \ac{rt} space for the flux and the discontinuous pressure space as described above.  This mixed \ac{fe} pair satisfies the inf-sup condition and, thus, it is stable.

\begin{minted}[bgcolor=bg]{julia1}
order = 1
V = TestFESpace(
  reffe=:RaviartThomas, order=order, valuetype=VectorValue{2,Float64},
  conformity=:HDiv, model=model, dirichlet_tags=[5,6])
Q = TestFESpace(
  reffe=:QLagrangian, order=order, valuetype=Float64,
  conformity=:L2, model=model)
\end{minted}
Note that the Dirichlet boundary for the flux are the bottom and top sides of the squared domain (identified with the boundary tags 5, and 6 respectively), whereas no Dirichlet data can be imposed on the pressure space. We select \shb{conformity=:HDiv} for the flux (i.e., shape functions with $H^1(\mathrm{div};\Omega)$ regularity) and \shb{conformity=:L2} for the pressure (i.e. discontinuous shape functions).

From these objects, we construct the  trial spaces. Note that we impose homogeneous boundary conditions for the flux.
\begin{minted}[bgcolor=bg]{julia1}
uD = VectorValue(0.0,0.0)
U = TrialFESpace(V,uD)
P = TrialFESpace(Q)
\end{minted}

When the singe-field spaces have been designed, the multi-field test and trial spaces are constructed from arrays of single-field ones in a natural way.
\begin{minted}[bgcolor=bg]{julia1}
Y = MultiFieldFESpace([V, Q])
X = MultiFieldFESpace([U, P])
\end{minted}

\subsection{Numerical integration}

In this example we need to integrate in the interior of $\Omega$ and on the Neumann boundary $\Gamma_{\rm N}$. For the volume integrals, we extract the triangulation from the geometrical model and define the corresponding cell-wise quadrature of degree of exactness at least 2 as follows.
\begin{minted}[bgcolor=bg]{julia1}
trian = Triangulation(model)
degree = 2
quad = CellQuadrature(trian,degree)
\end{minted}
In order to integrate the Neumann boundary condition, we only need to build an integration mesh for the right side of the domain (which is the only part of $\Gamma_{\rm N}$, where the Neumann function $h$ is different from zero). Within the model, the right side of $\Omega$ is identified with the boundary tag 8. Using this identifier, we extract the corresponding surface triangulation and create a quadrature with the desired degree of exactness.

\begin{minted}[bgcolor=bg]{julia1}
neumanntags = [8,]
btrian = BoundaryTriangulation(model,neumanntags)
bquad = CellQuadrature(btrian,degree)
\end{minted}

\subsection{Weak form}

We start by defining the permeability tensors commented above using the \shb{@law} macro.

\begin{minted}[bgcolor=bg]{julia1}
const kinv1 = TensorValue(1.0,0.0,0.0,1.0)
const kinv2 = TensorValue(100.0,90.0,90.0,100.0)
@law function σ(x,u)
   if ((abs(x[1]-0.5) <= 0.1) && (abs(x[2]-0.5) <= 0.1))
      return kinv2*u
   else
      return kinv1*u
   end
end
\end{minted}
With this definition, we can express the integrand of the bilinear form as follows. 
\begin{minted}[bgcolor=bg]{julia1}
px = get_physical_coordinate(trian)
function a(x,y)
   v, q = y
   u, p = x
   v*σ(px,u) - (∇*v)*p + q*(∇*u)
end
\end{minted}
The arguments \shb{x} and \shb{y} of previous function represent a trial and a test function in the multi-field trial and test spaces \shb{X} and \shb{Y} respectively. In the two first lines in the function definition, we unpack the single-field test and trial functions from the multi-field ones. E.g., \shb{v} represents a test function for the flux and \shb{q} for the pressure. These quantities can also be written as \shb{y[1]} and \shb{y[2]} respectively. From the single-field functions, we write the different terms of the bilinear form as we have done in previous tutorials.
In a similar way, we can define the forcing term related to the Neumann boundary condition.
\begin{minted}[bgcolor=bg]{julia1}
nb = get_normal_vector(btrian)
h = -1.0
function b_ΓN(y)
  v, q = y
  (v*nb)*h
end
\end{minted}

\subsection{Multi-field \ac{fe} problem}

\begin{figure}[ht!]
\includegraphics[width=0.6\textwidth]{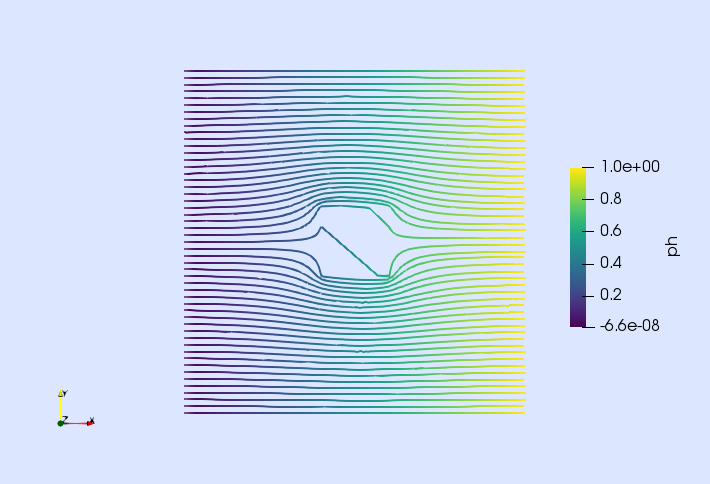}
\caption{\tutorial: View of the computed numerical solution. Streamlines of the flux colored by the value of the pressure.}
\label{fig:darcy_results}
\end{figure}

Finally, we can assemble the \ac{fe} problem and solve it. Note that we build the \shb{AffineFEOperator} object using the multi-field trial and test spaces \shb{X} and \shb{Y}.
\begin{minted}[bgcolor=bg]{julia1}
t_Ω = LinearFETerm(a,trian,quad)
t_ΓN = FESource(b_ΓN,btrian,bquad)
op = AffineFEOperator(X,Y,t_Ω,t_ΓN)
xh = solve(op)
uh, ph = xh
\end{minted}
Since this is a multi-field example, the \shb{solve} function returns a multi-field solution \shb{xh}, which can be unpacked in order to finally recover each field of the problem. The resulting single-field objects can be visualized as in previous tutorials (see \fig{fig:darcy_results}).
\begin{minted}[bgcolor=bg]{julia1}
writevtk(trian,"darcyresults",cellfields=["uh"=>uh,"ph"=>ph])
\end{minted}

\stepcounter{tutorial}
\section{Incompressible Navier-Stokes equations (\tutorial)} \label{sec:ins}

In this tutorial, we will learn
\begin{itemize}
\item How to solve nonlinear multi-field \acp{pde} in \gridap
\item How to build \ac{fe} spaces whose functions have zero mean value
\end{itemize}

\subsection{Problem statement}

The goal of this last tutorial is to solve a nonlinear multi-field \ac{pde}. As a model problem, we consider a well known benchmark in computational fluid dynamics,  the lid-driven cavity for the incompressible Navier-Stokes equations. Formally, the \ac{pde} we want to solve is: find the velocity vector $u$ and the pressure $p$ such that
\begin{align}
\left\lbrace
\begin{aligned}
-\Delta u + \mathit{Re}\ (u\cdot \nabla)\ u + \nabla p = 0 &\text{ in }\Omega,\\
\nabla\cdot u = 0 &\text{ in } \Omega,\\
u = g &\text{ on } \partial\Omega,
\end{aligned}
\right.
\end{align}
where the computational domain is the unit square $\Omega \doteq (0,1)^d$, $d=2$, $\mathit{Re}$ is the Reynolds number (here, we take $\mathit{Re}=10$), and $(w \cdot \nabla)\ u = (\nabla u)^t w$  is the well known convection operator. In this example, the driving force is the Dirichlet boundary velocity $g$, which is a non-zero horizontal velocity with a value of $g = (1,0)^t$ on the top side of the cavity, namely the boundary $(0,1)\times\{1\}$, and $g=0$ elsewhere on $\partial\Omega$. Since we impose Dirichlet boundary conditions on the entire boundary $\partial\Omega$, the mean value of the pressure is constrained to zero in order have a well posed problem,
\begin{equation}
\int_\Omega q \ {\rm d}\Omega = 0.
\end{equation}

\subsection{Numerical Scheme} For the numerical approximation, we chose a formulation based on inf-sub stable $Q_k/P_{k-1}$ elements with continuous velocities and discontinuous pressures (see, e.g., \cite{elman_finite_2005} for specific details). The interpolation spaces are defined as follows.  The velocity interpolation space is
\begin{equation}
V \doteq \{ v \in [C^0(\Omega)]^d:\ v|_T\in [Q_k(T)]^d \text{ for all } T\in\mathcal{T} \},
\end{equation}
where $T$ denotes an arbitrary cell of the \ac{fe} mesh $\mathcal{T}$, and $Q_k(T)$ is the local polynomial space in cell $T$ defined as the multi-variate polynomials in $T$ of order less or equal to $k$ in each spatial coordinate. Note that, this is the usual continuous vector-valued Lagrangian \ac{fe} space of order $k$ defined on a mesh of quadrilaterals or hexahedra.  On the other hand, the space for the pressure~is
\begin{equation}
\begin{aligned}
Q_0 &\doteq \{ q \in Q: \  \int_\Omega q \ {\rm d}\Omega = 0\}, \text{ with}\\
Q &\doteq \{ q \in L^2(\Omega):\ q|_T\in P_{k-1}(T) \text{ for all } T\in\mathcal{T}\},
\end{aligned}
\end{equation}
where $P_{k-1}(T)$ is the polynomial space of multi-variate polynomials in $T$ of degree less or equal to $k-1$. Note that functions in $Q_0$ are strongly constrained to have zero mean value. This is achieved in the code by removing one \ac{dof} from the (unconstrained) interpolation space $Q$ and  adding a constant to the computed pressure so that the resulting function has zero mean value.

The weak form associated to these interpolation spaces reads: find $(u,p)\in U_g \times Q_0$ such that $[r(u,p)](v,q)=0$ for all $(v,q)\in V_0 \times Q_0$
where $U_g$ and $V_0$ are the set of functions in $V$ fulfilling the Dirichlet boundary condition $g$ and $0$  on $\partial\Omega$ respectively. The weak residual $r$ evaluated at a given pair $(u,p)$ is the linear form defined as
\begin{equation}
[r(u,p)](v,q) \doteq a((u,p),(v,q))+ [c(u)](v),
\end{equation}
with 
\begin{align}
\begin{aligned}
a((u,p),(v,q)) &\doteq \int_{\Omega} \nabla v \cdot \nabla u \ {\rm d}\Omega - \int_{\Omega} (\nabla\cdot v) \ p \ {\rm d}\Omega + \int_{\Omega} q \ (\nabla \cdot u) \ {\rm d}\Omega,\\
[c(u)](v) &\doteq \int_{\Omega} v 	\cdot \left( (u\cdot\nabla)\ u \right)\ {\rm d}\Omega.\\
\end{aligned}
\end{align}
Note that the bilinear form $a$ is associated with the linear part of the \ac{pde}, whereas $c$ is the contribution to the residual resulting from the convective term.

In order to solve this nonlinear weak equation with a Newton-Raphson method, one needs to compute the Jacobian associated with the residual $r$. In this case, the Jacobian $j$ evaluated at a pair $(u,p)$ is the bilinear form defined as
\begin{equation}
[j(u,p)]((\delta u, \delta p),(v,q)) \doteq a((\delta u,\delta p),(v,q))  + [{\rm d}c(u)](v,\delta u),
\end{equation}
where ${\rm d}c$ results from the linearization of the convective term, namely
\begin{equation}
[{\rm d}c(u)](\delta u, v) \doteq \int_{\Omega} v \cdot \left( (u\cdot\nabla)\ \delta u \right) \ {\rm d}\Omega + \int_{\Omega} v \cdot \left( (\delta u\cdot\nabla)\ u \right)  \ {\rm d}\Omega. 
\end{equation}

The implementation of this numerical scheme is done in \gridap{} by combining the concepts previously seen for single-field nonlinear \acp{pde}  and linear multi-field problems.

\subsection{Discrete model}

We start with the discretization of the computational domain. We consider a $100\times100$ Cartesian mesh of the unit square.
\begin{minted}[bgcolor=bg]{julia1}
using Gridap
n = 100
domain = (0,1,0,1)
partition = (n,n)
model = CartesianDiscreteModel(domain,partition)
\end{minted}

For convenience, we create two new boundary tags,  namely \shb{"diri1"} and \shb{"diri0"}, one for the top side of the square (where the velocity is non-zero), and another for the rest of the boundary (where the velocity is zero).
\begin{minted}[bgcolor=bg]{julia1}
labels = get_face_labeling(model)
add_tag_from_tags!(labels,"diri1",[6,])
add_tag_from_tags!(labels,"diri0",[1,2,3,4,5,7,8])
\end{minted}

\subsection{\ac{fe} spaces} For the velocities, we need to create a conventional vector-valued continuous Lagrangian \ac{fe} space. In this example, we select a second order interpolation.
\begin{minted}[bgcolor=bg]{julia1}
D = 2
order = 2
V = TestFESpace(
  reffe=:Lagrangian, conformity=:H1, valuetype=VectorValue{D,Float64},
  model=model, labels=labels, order=order, dirichlet_tags=["diri0","diri1"])
\end{minted}
The interpolation space for the pressure is built as follows
\begin{minted}[bgcolor=bg]{julia1}
Q = TestFESpace(
  reffe=:PLagrangian, conformity=:L2, valuetype=Float64,
  model=model, order=order-1, constraint=:zeromean)
\end{minted}
With the options \shb{reffe=:PLagrangian}, \shb{valuetype=Float64}, and \shb{order=order-1}, we select the local polynomial space $P_{k-1}(T)$ on the cells $T\in\mathcal{T}$. With the symbol \shb{:PLagrangian} we specifically chose a local Lagrangian interpolation of type ``P". Using \shb{:Lagrangian}, would lead to a local Lagrangian of type ``Q" since this is the default for quadrilateral or hexahedral elements. On the other hand, \shb{constraint=:zeromean} leads to a \ac{fe} space, whose functions are constrained to have mean value equal to zero as required in this example. With these objects, we build the trial and multi-field \ac{fe} spaces
\begin{minted}[bgcolor=bg]{julia1}
uD0 = VectorValue(0,0)
uD1 = VectorValue(1,0)
U = TrialFESpace(V,[uD0,uD1])
P = TrialFESpace(Q)

Y = MultiFieldFESpace([V, Q])
X = MultiFieldFESpace([U, P])
\end{minted}

\subsection{Nonlinear weak form} The different terms of the nonlinear weak form for this example are defined following an approach similar to the one discussed for the $p$-Laplacian equation, but this time using the notation for multi-field problems.
\begin{minted}[bgcolor=bg]{julia1}
const Re = 10.0
@law conv(u,∇u) = Re*(∇u')*u
@law dconv(du,∇du,u,∇u) = conv(u,∇du)+conv(du,∇u)

function a(x,y)
  u, p = x
  v, q = y
  inner(∇(v),∇(u)) - (∇*v)*p + q*(∇*u)
end

c(u,v) = v*conv(u,∇(u))
dc(u,du,v) = v*dconv(du,∇(du),u,∇(u))

function res(x,y)
  u, p = x
  v, q = y
  a(x,y) + c(u,v)
end

function jac(x,dx,y)
  u, p = x
  v, q = y
  du, dp = dx
  a(dx,y)+ dc(u,du,v)
end
\end{minted}

With the functions \shb{res}, and \shb{jac} representing the weak residual and the Jacobian, we build the nonlinear \ac{fe} problem:
\begin{minted}[bgcolor=bg]{julia1}
trian = Triangulation(model)
degree = (order-1)*2
quad = CellQuadrature(trian,degree)
t_Ω = FETerm(res,jac,trian,quad)
op = FEOperator(X,Y,t_Ω)
\end{minted}

\subsection{Nonlinear solver phase}

To finally solve the problem, we consider the same nonlinear solver as previously considered for the  $p$-Laplacian equation.
\begin{minted}[bgcolor=bg]{julia1}
using LineSearches: BackTracking
nls = NLSolver(
  show_trace=true, method=:newton, linesearch=BackTracking())
solver = FESolver(nls)
\end{minted}

In this example, we solve the problem without providing an initial guess (a default one equal to zero will be generated internally)
\begin{minted}[bgcolor=bg]{julia1}
uh, ph = solve(solver,op)
\end{minted}

\begin{figure}[ht!]
\includegraphics[width=0.6\textwidth]{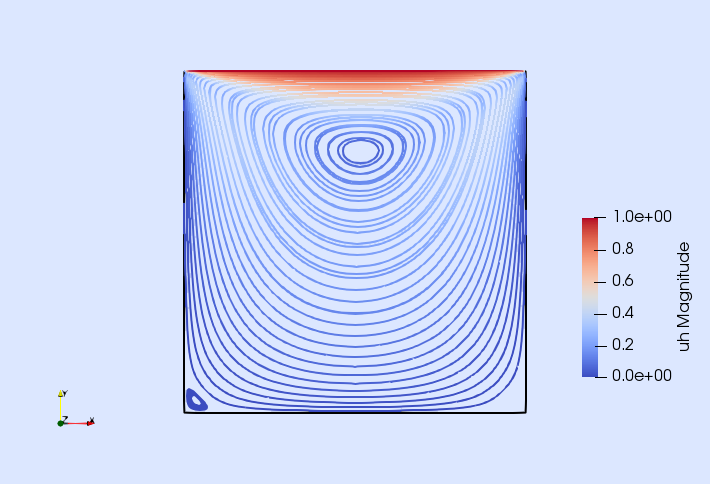}
\caption{\tutorial: Streamlines of the computed velocity field colored by the velocity magnitude.}
\label{fig:ins_solution}
\end{figure}

Finally, we write the results for visualization (see \fig{fig:ins_solution}).
\begin{minted}[bgcolor=bg]{julia1}
writevtk(trian,"ins-results",cellfields=["uh"=>uh,"ph"=>ph])
\end{minted}

\section{Conclusions}\label{sec:con}

In this work, we have presented a tutorial-driven introduction to \gridap{}. We have covered some common \ac{pde} problems and their discretization with conforming \acp{fe} and \ac{dg}. The \gridap{} framework is very expressive, allowing the implementation of complex problems in a reduced number of code lines, using at a very high level the abstractions provided by the library (e.g., discrete models, \ac{fe} spaces, triangulations, quadratures, etc.). The implementation of weak formulations of \acp{pde} highly resembles the {LaTeX} form, boosting productivity. In any case, this can also be attained using dynamic languages as Python. What makes \gridap{} special is that we want to provide such productive environment without sacrificing performance or expressiveness. This is achieved by writing \gridap{} exclusively in \julia{} and with a careful  design to minimize dynamic memory allocations, keep type-stability, and providing the \ac{jit} system enough information to infer types.

The design of \gridap{} is quite unique for a numerical \ac{pde} software, since it does not follow neither procedural nor object-oriented paradigms. Instead, we exploit multiple dispatching and functional programming capabilities provided by Julia. Most of the structures in \gridap{} are immutable and based on lazy containers to represent cell-wise data. 
A detailed exposition of the design patterns of \gridap{} and its performance analysis will be published elsewhere.

\gridap{} is a young project and there are still many things to be done. In particular, we want to provide in the near future Julia native shared and (possibly) distributed memory parallelization, MPI-based parallelization, support for adaptive mesh refinement and distributed octree meshes, and wrappers for popular solver libraries.

\section*{Acknowledgments}

SB gratefully acknowledges the support received from the Catalan Government through the ICREA Acad\`emia Research Program. FV gratefully acknowledges the support received from the Secretaria d’Universitats i Recerca of the Catalan Government in the framework of the Beatriu Pinós Program (Grant Id.: 2016 BP 00145).

\setlength{\bibsep}{0.0ex plus 0.00ex}
\bibliographystyle{myabbrvnat}
\bibliography{art040}

\end{document}